\documentclass{aa}  

\usepackage{graphicx}
\usepackage{placeins}
\usepackage{txfonts}
\usepackage{ulem}
\usepackage{pdflscape}
\usepackage{amsmath}
\usepackage[caption=false]{subfig}
\usepackage{hyperref}

\usepackage{mathabx}
\usepackage{hyperref}
\usepackage{tablefootnote}

\hypersetup{colorlinks,allcolors=blue}

\newcommand{\MEarth}{M$_{\mathrm{\oplus}}$\,}
\newcommand{\MJup}{M$_{\mathrm{Jup}}$\,}

\newcommand{\RSun}{R$_{\odot}$\,}
\newcommand{\Msun}{M$_{\odot}$\,}
\newcommand{\Mjup}{M$_{\rm Jup}$\,}

\newcommand{\teff}{$T_{\rm eff}$\,}

\newcommand{\mdot}{$\dot{M}$\,}

\def\tablebib#1{\par\vspace*{2ex}%
 \parbox{\hsize}{\leftskip0pt\rightskip0pt
 {\noindent\small\textbf{References.}\, #1\par}}}
\begin{document}

   \title{The evolution of discs and the epoch of formation of giant planets around solar type stars}


   \author{Raffaele Gratton\inst{1}, Elisabetta Rigliaco\inst{1} }

   \institute{INAF – Osservatorio Astronomico di Padova, Vicolo dell’Osservatorio 5, 35122 Padova, Italy}


   \date{Received 27 February 2026 / Accepted 15 September 2026 }


\abstract{Disc evolution dictates the time for planet formation providing the raw material to build the planets' cores. Dynamical stellar masses derived from the kinematics of the gas in discs provide constraints on the ages of very young stars, offering a powerful approach to studying disc evolution and the epoch of giant-planet formation. We investigated a sample of 33 solar-type stars with bright protoplanetary discs and ages between 1 and 9\,Myr. We modeled the spectral energy distributions and ALMA dust continuum images simultaneously to constrain the stellar radius and effective temperature, accretion rate, disc dust mass, the radii of the inner and outer disc regions (roughly separated at the ice line), and the maximum grain size in the inner disc. Combining the stellar masses and radii with stellar evolutionary models, we derived individual stellar ages. We found that the accretion rate and the dust mass of the inner disc evolve with time. The dust mass of the inner disc correlates with the observed accretion rates for stars younger than 2\, Myr. A depletion of centimetre-sized grains in the inner disc is observed in more than half of the analysed stars with ages between 2 and 5\, Myr indicating the action of a mechanism that inhibits the inward drift of large grains, potentially associated with the progressive growth of giant-planet cores more massive than 30 \MEarth, although the exact mass threshold remains uncertain. In addition, three discs exhibiting strong asymmetries in their ALMA continuum emission are associated with stars with ages of $\sim$2\, Myr. The results provide observational indications of significant changes in the dust distribution during the epoch when giant planets may be forming.}


\titlerunning{The evolution of discs around solar type stars }
\authorrunning{R. Gratton and E. Rigliaco}

   \maketitle
%

\section{Introduction}
\label{sect:Introduction}

Planets are thought to form within the circumstellar discs that mediate the accretion of material from the interstellar medium onto the stars. Two main mechanisms for planet formation have been proposed: disc instability \citep{Boss1997, Kratter2016} and core accretion \citep{Safronov1972, Pollack1996, Mordasini2012, Armitage2020}. The first mechanism is very fast with typical timescales of a few dynamical times, corresponding to a few tens of thousands of years or less. The second mechanism is much more complex and slow, with a typical timescale of a few million years. For many reasons, the core accretion scenario (and its variant the pebble accretion scenario) is thought to be more appropriate for describing the formation of the Solar System and for most exoplanets \citep{Lambrechts2012, Lambrechts2014,  Bitsch2015, Ormel2026}. Since discs also evolve on timescales of a few million years and the formation of planets also contributes to the evolution and dispersal of the disc, there is a complex interplay between planet formation and disc evolution.

A specific argument of interest is the timescale of the formation of giant planets. For what concerns the Solar System, important information is provided by the study of the isotopic ratios in meteorites, especially those concerning unstable isotopes with lifetimes of the order of a million years. Meteorites have largely been used to date the early phases of formation of the Solar System. A discussion of current understanding can be found in \citet{Kruijer2017}. Current evidence suggests that a first important differentiation between the properties of the inner and outer parts of the proto-solar disc occurred about a million years after beginning, with the separation between non-carbonaceous chondrites (NC, in the inner Solar System) and carbonaceous chondrites (CC, in the outer Solar System). The formation of Jupiter's core may have played a role acting as a filter that stops the inward drift of some type of dust grains. However, the final growth of Jupiter likely required several more million years \citep{Alibert2018}.

How the scenario for the Solar System compares with observations of young exoplanetary systems is not yet entirely clear. One of the main reasons is the lack of accurate enough ages for individual very young stars. In addition, there are clear trends with stellar mass in the properties of exoplanets, namely concerning the frequency of giant planets, which are very rare around low-mass stars \citep{Bryant2023}. Although this fact has often been used to constrain models for giant planet formation (e.g., \citealt{Ida2005, Burn2024}), the models themselves strongly depend on properties of the discs that are not well known. Accurate ages for a large enough set of very young stars with discs would allow a better understanding of the correct timescales involved.

Several previous studies examined the characteristics of discs in a systematic way. They mainly focused on systematic trends of discs with stellar mass (e.g., \citealt{Pascucci2016, Ansdell2016}) or taxonomy, trying to group objects with similar morphologies (e.g., \citealt{Garufi2018, Garufi2024, Parker2022}). We rather focus on the temporal evolution of some basic properties of discs (dust masses, radii, grain content) for stars with masses similar to the Sun. Dynamical masses are now becoming available for a conspicuous number of very young stars from observation of molecular gas with the Atacama Large Millimeter/submillimeter Array (ALMA). Modelling the kinematics of the gas allowed several authors to determine the mass of the central object. Once reliable masses are available, ages may be derived from the stellar radii using a comparison with evolutionary models. In this paper, we will exploit this fact to determine accurate ages of a sample of 33 solar type disc hosting stars (masses in the range from 0.7 to 1.4 \Msun). Employing a simple code that simultaneously fits the spectral energy distribution (SED) and high resolution images of dust continuum emission from ALMA, we also derived basic parameters of the disc and the accretion rate on the star uniformly throughout this sample. We used these results to discuss the evolution of discs and the epoch of formation of giant planets around stars similar to the Sun. 

We present the data used in our study in Sect. \ref{sect:data}. The model used in our analysis is described in Sect. \ref{sect:methods}. The results are presented in Sect. \ref{sect:results} and discussed in Sect. \ref{sect:discussion}. Conclusions are drawn in Sect. \ref{sect:conclusions}.

\section{Data}
\label{sect:data}

In this paper we discuss the evolution of protoplanetary discs around the solar type stars listed in Table\, \ref{tab:uv_data}. We used both the SED, which provides information about the central region of the system, and ALMA band 6 or 7 continuum images, which yield data about the outer regions of the disc. ALMA data at high spatial resolution are important in our analysis because they allow us to constrain the emission at long wavelengths from the inner regions of the disc, roughly within the ice line, which is among the most important results of our study. Since these two sets of data cannot be considered as completely independent, we performed a simultaneous fit of them using a simple but physically informed parametric model. 

The parameters that we derived using this approach, listed in Table\, \ref{tab:range}, can be placed in an evolutionary context using accurate ages for individual stars that may be derived from a comparison with isochrones for stars whose dynamical masses $M_\star$ are obtained from the kinematics of the disc. 
In particular, we considered ages derived from the stellar radii because the radius is rapidly changing during the pre-main sequence evolution. Hereafter, we present the dataset we used.

\subsection{The sample}
\label{sect:sample}

We first selected stars with masses determined from disc kinematics traced by molecular line observations obtained with ALMA.
We considered stars with dynamical masses determined in various papers: \citet{Simon2019}, \citet{Keppler2019}, \citet{Braun2021}, \citet{Bosman2021}, \citet{Miley2024}, \citet{Trapman2025}, \citet{Izquierdo2025}, and \citet{Longarini2025}. We also considered the list of young stars with protoplanetary disc compiled by Olga Eretnova and Sergei Khaibrakhmanov\footnote{\url{https://zenodo.org/records/17853557}} but we did not find there additional objects with the required data. Whenever a star was considered in more samples, we adopted the value of the mass from the most recent paper. The sample only includes stars with masses $M_\star$ in the range between 0.7 and 1.4 \, \Msun. Stars of higher mass were excluded because their evolution is so fast that we cannot derive reliable ages with our approach. Stars of lower mass were excluded because the evolution of their disc is likely different from that of solar type stars as suggested by the very low frequency of giant planets around them \citep{Kennedy2008, Johnson2010, Morales2019}, and will be discussed in a following paper. 

Given the limitations in the model we used in our analysis (see Sect. \ref{sect:methods}), we only considered stars whose disc is seen with an inclination $i<70$\, deg. We also discarded stars that have close bright companions within about 2 arcsec (about 300 au at the distance of the programme stars) with separate discs around them. Finally, we excluded three stars (Haro 6--13, 2MASS J16070854--3914075, and CIDA 9A aka IRAS 05022+2527) that are still completely embedded in the dust, for which our model is too simplistic.

The final sample counts 33 stars with an average mass of 0.96\, \Msun (see Tables \ref{tab:uv_data} and  \ref{tab:input_data}). Therefore, the sample may be used to consider the evolution of discs around the solar type stars. However, we should keep in mind that, by construction, our sample is biassed towards long living discs. Many stars as old as those considered in this paper do not have any more discs that are bright enough for the determination of their mass. In addition, this sample is drawn from local star forming regions (Taurus, Lupus, Upper Scorpius, Coronae Australis, Chamaeleontis) that typically have low density and may then be more favourable environments for the formation of giant planets \citep{Gratton2024, Gratton2025}.

Undisputed evidence for planetary companions was found for two of the sample stars: PDS\, 70 \citep{Keppler2018} and 2MASS J16120668--3010270 \citep{Li2025}. The (giant) planets are also observed in H$\alpha$ in both systems \citep{Haffert2019, Li2025}, showing that they are accreting objects. Candidate planets have been proposed for a number of other stars in our sample. A suspected protoplanet at 100 au with a mass in the 2-3~\Mjup range was proposed near IM Lup \citep{Verrios2022}. Possible planets at 2.4\, au from GK Tau and about 20\,au from CQ Tau were suggested by \citet{Marsh1993}. Possible protoplanets have been proposed around LkCa 15 \citep{Sallum2015,Close2025} but no consensus has yet been reached (e.g. \citealt{Currie2019, Huelamo2022}). \citet{Rigliaco2026} found some evidence for a planet of $M<3$\,\Mjup at 25.7\,au from WRAY15--1880 aka RX\,J184257--35327. Finally, \citet{Ribas2025} proposed a planet at 1--3\,au from PDS 66, and very recently \citet{Yoshida2026} proposed a second one, at about 60\, au from this same star.

We also noticed that some of the stars were observed with high contrast imaging in H$\alpha$ without detection of possible companions. They include MWC 758 \citep{Cugno2019, Zurlo2020}, PDS 66 \citep{Zurlo2020}, WRAY~15--1880 \citep{Rigliaco2026}, LkCa~15, and RX\,J1615.3--3255 \citep{Huelamo2022}

Finally, we used our approach to estimate the age of TYC 5709-354-1, which is known to host an accreting giant planet \citep{Close2025}, although we lack publicly available high-resolution ALMA images for this object. For this star, we adopted a mass of 1.080 \Msun from \citet{vanCapelleveen2025} and simply fitted the SED to derive the radius. We obtained an age of $6.1\pm 1.0$\, Myr. TYC 5709-354-1 is included here solely to place the observation of accreting planets within the context of our study.

\subsection{Spectral energy distributions}

The SED of the 33 stars were obtained using the values provided by the SIMBAD VizieR Photometry viewer \citep{Ochsenbein2000}\footnote{\url{https://cds.unistra.fr/news/2014/11/13-photometry-viewer/}}. We generally gave preference to values provided by {\it Gaia} DR3 \citep{GaiaCollaboration2022}, 2MASS \citep{Skrutskie2006}, {\it WISE} (WISE All-Sky Data Release, \citealt{Wright2010, Cutri2012}), {\it Spitzer} \citep{Spitzer2021}, {\it Herschel} (PACS: \citealt{Ribas2017};  Herschel/SPIRE point source catalogue (HSPSC): \citealt{Herschel2024}), Submillimetre Common-User Bolometer Array (SCUBA) at the James Clerk Maxwell Telescope  \citep{Mohanty2013}, and ALMA, but also considered additional near UV and $B$-band photometry from the SkyMapper Southern Sky Survey DR1.1 \citep{Wolf2018}. 

Special care was devoted to gathering space UV data that are crucial to obtain sensible values for the accretion rates (see Table \ref{tab:uv_data}). UV data were obtained from an inspection of data obtained using the Cosmic Origins Spectrograph (COS) and the Space Telescope Imaging Spectrograph (STIS) on board of the {\it Hubble Space Telescope} as available in the Barbara A. Mikulski Archive\footnote{\url{https://mast.stsci.edu/portal/Mashup/Clients/Mast/Portal.html}}. In addition, we considered data from International Ultraviolet Explorer (IUE) \citep{Boggess1978, Beitia-Antero2016}, Galaxy Evolution Explorer (GALEX) \citep{Bianchi2011}, and {\it XMM-Newton} \citep{Page2023}. Overall, space UV data are available for all but five stars (2MASS J16202863--2442087, 2MASS J16120668--3010270, V1094 Sco, 2MASS J16090075--1908526, RX\, J1604.3--2130A). They are among the oldest in our sample and are likely faint in the UV. Tables listing all the photometric points for building the SED shown in the figures are available in the Vizier database at the Centre de Données astronomiques de Strasbourg (CDS).

The photometry was corrected for interstellar absorption using the law by \citet{Cardelli1989} assuming a ratio of $R_{V}=3.1$\, mag between total and selective absorption. The value of the total interstellar absorption in the $V$ band for each star was obtained from the difference between the observed $B_P-R_P$ colour in {\it Gaia} DR3 and that expected for stars of the same spectral type according to the tables by \citet{Pecaut2013}\footnote{\url{https://www.pas.rochester.edu/\, emamajek/EEM_dwarf_UBVIJHK_colors_Teff.txt}}, assuming $E(B_P-R_P)=0.457\,A_V$ as given by the law of \citet{Cardelli1989}.

\subsection{ALMA dust images}

We searched in the ALMA Science archive \footnote{\url{https://almascience.eso.org/aq/}} for high resolution data in bands 6 and 7 ($\rho<0.15$\, arcsec, corresponding to less than 20~au at the typical distance of the targets). We considered data publicly available reduced with the {\tt CASA} software\footnote{\url{https://casa.nrao.edu/index.shtml}}. When more datasets were available, we gave preference to those providing the highest resolution and lowest noise. Such public data are not available for all stars whose dynamical masses were available, and this limits the number of stars included in our sample.

Whenever a dataset was available, we downloaded the reduced continuum images, considering the latest reduction. The datasets used in the analysis are listed in Table \ref{tab:alma_data}. The version of the {\tt CASA} software used in the reduction is given in Column 4. Continuum sensitivity (Cont. sens., Column 5) and resolution (res., Column 6) in the table are as given in the ALMA science archive \footnote{\url{https://almascience.eso.org/aq/}}. Archived images are in physical flux units and have a variable scale and resolution, as defined by the beam parameters contained in the header of the fits files, and used in our analysis.

\section{Methods}
\label{sect:methods}

\subsection{Model description}

We derived stellar and disc main properties (radius, effective temperature, accretion rate, dust mass and radius of the inner and outer regions of the disc, and maximum radius of the grains in the inner disc) by fitting simultaneously the SED and ALMA high-resolution images of the continuum millimetre emission. The model was kept as simple as possible (e.g., for what concerns the properties of the grains and geometry of the disc) to reduce the number of free parameters, allowing a simpler comparison between the properties of the various discs, although the simplifying assumptions introduce some ambiguity in the physical interpretation of individual model parameters.

In the proposed model, the SED is computed for 300 wavelengths values ($\lambda$ in $\mu$m) determined with the following law: $\lambda=0.01+0.001*j^{2.5}$, with $0<j<299$. It includes emission by three different regions, as listed below: the stellar photosphere, the accretion region, and the dusty disc. We assumed that gas did not directly contribute to the SED and to the continuum disc emission, but it contributed to the accretion luminosity. We neglected the contribution by light reflected on the disc surface because this is much smaller than the thermal emission of the inner portions of the disc. 
\begin{itemize}
    \item {Emission of the stellar photosphere.} The emission by the photosphere was simulated with model atmospheres extracted from the Kurucz grid\footnote{\url{https://www.stsci.edu/hst/instrumentation/reference-data-for-calibration-and-tools/astronomical-catalogs/kurucz-1993-models}} assuming a surface gravity of $\log{g}=4.0$, matching the sample stars and a solar metal abundance. Therefore, it was characterised by the stellar effective temperature \teff and radius $R_*$\, which are free parameters in the fit. We did not consider any contribution from the chromosphere, that might represent a significant contribution to the UV excess for the oldest stars in our sample.

    \item{Emission of the accretion region.} Accretion onto the star causes a shock that heats a fraction of the covered stellar surface \citep{Calvet1998}. The emission from the region heated by the shock front was taken into account as a black body with a fixed temperature of 10,000\, K (representing a typical value, see \citealt{Pittman2025}) and a variable area; the accretion luminosity was derived from the mass accretion rate \mdot considering the stellar radius and mass. We assumed that half of the contraction energy is radiated (as expected for viscous discs) and that the ratio between the inner edge (of the gas) disc and the star radius is 4, as in \citet{Alcala2017}. \mdot was a free parameter in the fit. Since we adopted a black body emission at a constant temperature of 10,000 K for the accretion region, the area of the stellar surface covered by the accretion region simply results from the accretion luminosity, which in turn is directly related to the accretion rate (via mass and radius of the star). Typical values yield a fraction of the stellar surface that is $\leq 0.002$.

    \item{Dust emission from the disc.} We assumed that disc emission is only thermal emission from dust, neglecting the possible contribution of other mechanisms such as gas free-free emission, which may be important at radio-wavelengths \citep{Rota2024, Garufi2025}. We calculated disc emission assuming a bi-dimensional and radially symmetric distribution of grains. We considered the area and local temperature for each of 500 logarithmically distributed annuli of the disc, taking into account self-absorption. The adopted approximation neglects the vertical structure of the discs and is not adequate for discs seen at a high inclination ($i>70$\, deg), which were not considered in this study. The disc was assumed to be composed of dust grains and to consist of two regions: inner and outer discs, each characterised by an inner and outer radius. For each separation $a$ from the star, the temperature was the equilibrium temperature of the grains. The equilibrium temperature $T$ for any distance $a$ from the star was calculated using the following equation:
 \begin{gather}
    T=T_{\rm eff}\, \sqrt{s\, k\, R_*/2\, a}\, (1-A)^{0.25},
    \end{gather}
 where \teff and $R_*$ are the effective temperature and radius of the star, the factor $s=1+L_{\rm acc}/L_{\rm phot}$ takes into account the contribution of accretion to the luminosity, $A$\ is the Bond albedo of the grains, and $k$\ a suitable constant to take into account the various units. 
 It should be noted that the temperature profile considered here implies a grazing angle of incident starlight that is independent of distance (see eq. (1) of \citealt{Chiang1997}), in contrast to the conventional definition of a flat disc, for which the grazing angle scales with distance. We found that using a flat disc (eq. (4) of \citealt{Chiang1997}) leads to significantly cooler outer discs and poorer agreement with the observations, requiring larger values of $\chi^2$ than those obtained in our adopted model. In our model, the dust properties enter only through a constant albedo. In contrast, radiative transfer calculations predict dust temperatures that depend on the dust composition, size distribution, and total dust mass. Our treatment therefore represents a simplified description of the dust physics. Furthermore, we neglected the screening effect of the inner parts of the disc and the vertical structure of the disc. Hence, these temperatures may be overestimated, especially in thick (younger) discs.
\end{itemize}

We remark here that the observables we considered (thermal continuum emission by dust at optical, infrared, and sub-millimetre wavelengths) do not allow us to constrain the mass in gas as well as in planetesimals within the disc. Additional observables such as the emission in atomic and molecular lines should be used to constrain the mass in gas (see, e.g. \citealt{Carmona2014, Miotello2016, Woitke2019, Trapman2025}) and there are no available data that can usefully constrain the mass in planetesimals. Notwithstanding this, we prefer to use the mass of the disc as one of the parameters that describe the discs because this allows a simpler treatment of the properties of dust as a function of the temperature along the disc (see Appendix \ref{Sect:equations}).

The grains in the disc are assumed to have a power-law distribution with radius $R$ between some minimum and maximum values. We used a logarithmic scale for the grain radii. After some tests, we considered a fixed value of the exponent of $-2.81$ to reduce the number of free parameters in the model. This exponent is consistent with the average of the values considered by \citet{Woitke2019}, who used a linear scale for the radii. Since the exponent is higher than $-$3.0, this model assumes that grains with large radii (in the grain size interval  10$^{-4}-3$~cm) contribute most of the mass. We allowed for different values of the maximum radius for the grains in the inner ($R_{\rm max,grain,inner}$) and outer ($R_{\rm max,grain,outer}$) portions of the disc. Given the range of wavelengths considered, the minimum grain radius was always set at 0.1\, $\mu$m and the maximum radius was initially set at 3\,cm, but modified downward for the inner disc to obtain a good fit. We considered two species of grains: silicates and ices, with densities of 3.5\, g\,cm$^{-3}$ and 0.92\, g\,cm$^{-3}$, respectively \citep{Kataoka2014}. We assumed typical fractional masses of 0.00314 for silicates and 0.00686 for ices \citep{Miyake1993} adapting the original values given in that paper to produce a gas-to-dust ratio of 100 for the outer disc, which is the value typically used, and a Bond albedo of $A=0.15$ for both types of grains \citep[e.g.][]{Ueda2023}.  
In our model, silicates and ice grains can only survive where the equilibrium temperature is lower than their sublimation temperature, which we assumed to be 1500 K for silicates and 170 K for ices. In Appendix \ref{Sect:equations} we give more details on the relations we considered between the gas and dust masses of the discs and the surface density. In our approach, the gas-to-dust ratio depends on the presence of ice grains (in addition to silicate grains) that are only present beyond the ice line and on $R_{\rm max,grain}$, which is a free parameter. The gas-to-dust ratio is constant at 100 in the outer disc, which always is beyond the ice line. It is at least 318 within the ice line, but it may be much higher depending on the value of $R_{\rm max,grain,inner}$ in the inner disc.  

The grain opacity required to derive the emissivity and optical thickness of the disc was calculated from  equation (54) of \citet{Mordasini2014}. More in detail, we assumed that the extinction coefficient $Q$ that gives the ratio of the extinction cross section of a grain of radius $R$ for radiation of wavelength $\lambda$ to its geometric cross section $\pi R^2$ is:
\begin{gather}
{\rm if} \,x<0.375 \,{\rm then}\,Q=0.3\, x \nonumber\\
{\rm if} \,x>0.375 \,{\rm and}\, x<2.188 \,{\rm then}\,  Q=0.8\, x^2\nonumber\\
{\rm if} \,x>2.188 \,{\rm and}\, x<1000 \,{\rm then}\,  Q=2.0+4.0 x^{-1}\nonumber\\
{\rm if} \,x>1000 \,{\rm then}\, Q=2.0
\end{gather}
\noindent
where $x=2\, \pi\, R/\lambda$. Using these formulas, the opacity depends only on the ratio between the wavelength and the grain size, without any dependence on the chemical composition and structure of the grains, representing a rough approximation. More accurate values should in principle consider the composition of the grains and their structure (fluffy vs. compact grains), but this would introduce other parameters to the fit. The implicit assumption in our discussion is that the dependence of chemical composition and structure of the grains on position within the disc and age is absorbed into the only parameter that we left free (namely the distribution of grain sizes).

The inner and outer radii of the inner disc $a_{\rm min}$(inner), $a_{\rm max}$(inner), as well as the total (gas and dust) mass of the inner and outer discs $M_{\rm inner}$ and $M_{\rm outer}$, are free parameters in our model. The value of $a_{\rm min}$(inner) was forced to be higher than or equal to the distance to the star where the temperature is equal to that of sublimation of the silicates. The value of $a_{\rm min}$(outer) was forced to be higher than or equal to the radius of condensation of the ices that we assumed to be where the equilibrium temperature is 170\, K. The surface density profile of the inner disc is assumed to run as $1/a$; this implies an equal contribution to the mass of the inner disc by annuli of constant radial thickness. For a viscous disc, this corresponds to a constant inward velocity of the grains. The surface density profile of the outer disc was assumed to be proportional to the intensity of the emission along the major axis measured by ALMA; therefore, the inner and outer radii are simply the region where this quantity could be measured. 

We recognise that the adopted prescriptions for the disc structure and grain properties are approximate. However, our objective is to construct a model with the smallest possible number of free parameters, and this necessarily entails a degree of simplification. For example, the diversity of grain emissivities is represented by a single parameter, namely the maximum grain size, allowing us to easily compare different discs without the complication of a large number of parameters. However, some of the properties we derived for the discs (for instance, the dust masses) should be considered with caution and only valid within these approximations. However, since the same assumptions were made for all cases, the trends with age may still reveal relevant features of the disc evolution.

\subsection{Considerations about noise}
\label{sect:noise}

\begin{figure*}
\sidecaption
    \includegraphics[width=12cm]{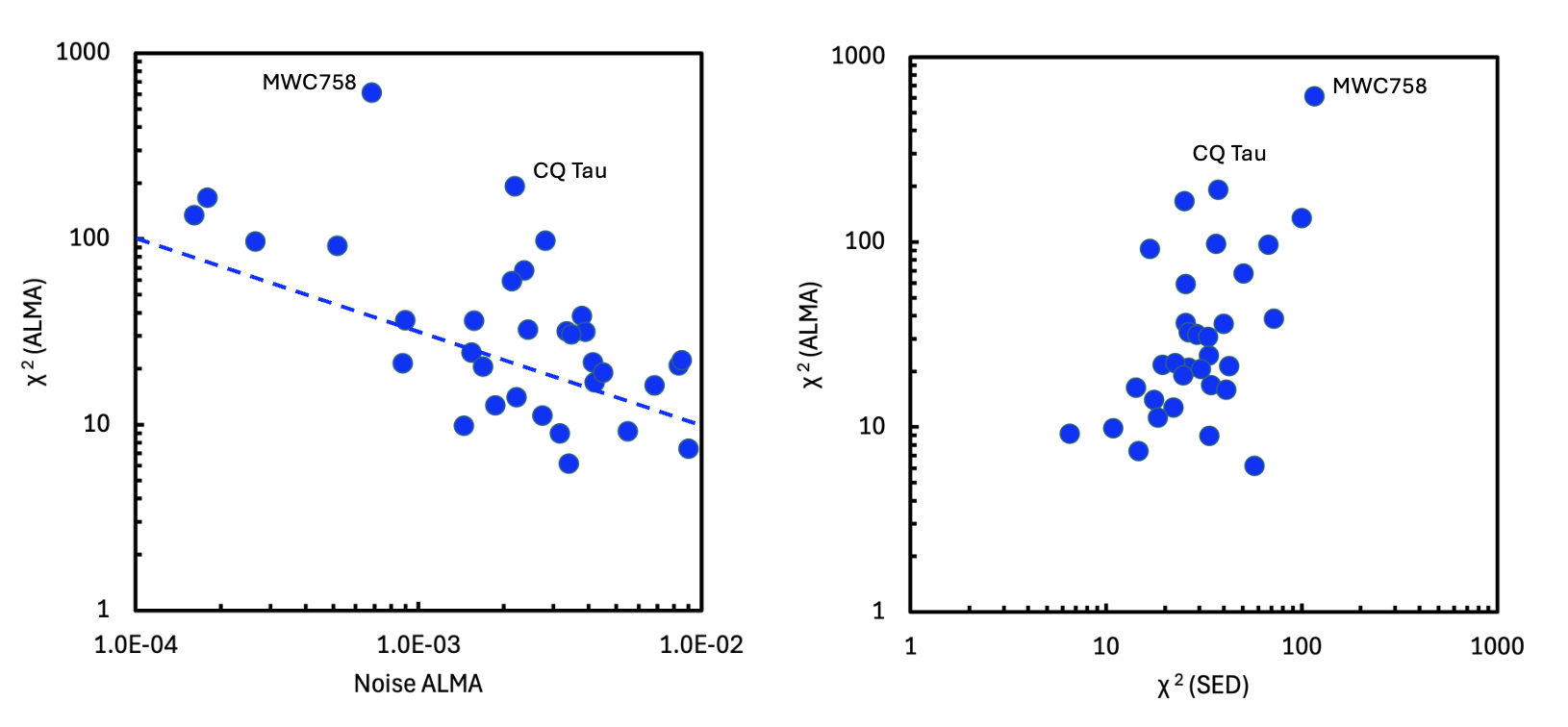}
    \caption{Left: $\chi^2$ obtained for the best fit of the ALMA images as a function of the noise in these images. Dashed line corresponds to the assumption that actual noise depends on the square root of the signal, as expected for photon noise. Right: $\chi^2$ obtained for the best fit of the ALMA images as a function of $\chi^2$ for the SED. }
    \label{fig:chisquare}
\end{figure*}

Keeping in mind the qualitative nature of the analysis of the results, and the simplification adopted in this analysis to reduce the free-parameters, we comment here how we treated the noise in our study. The noise in the photometric points considered in our analysis of the SED is the maximum between the error provided by the original source and an arbitrary value of 5\%. Since the sources may be variable, this avoids attributing too much weight to some of the photometric points.

The noise in ALMA images was obtained from the standard deviation in regions far from the star. Figure \ref{fig:chisquare} compares the $\chi^2$ obtained for the best fit of the ALMA images with the noise in these images and with the $\chi^2$ obtained for the SED. The first diagram shows that the lowest the noise in the ALMA images, the highest is the $\chi^2$ we obtain for the best fit, suggesting that our procedure underestimates the variance in the ALMA images, likely because the noise is not independent of the signal. 
The dashed line corresponds to the assumption that the noise scales with the square root of the signal, as expected for photon noise. The relation provides a good representation of most of the data, although some sources yield substantially higher $\chi^2$ values. The larger $\chi^2$ values may arise because some of the discs are not represented by a radially symmetric distribution of the emission, the most clear examples being MWC 758 and CQ Tau. Anyway, we think that the adoption of a noise independent of the signal yields a more appropriate overall fit because this choice avoids attributing too much weight to the large number of pixels with low signal with respect to the few ones with high signal. It effectively corresponds to the assumption that the weight of individual pixels is proportional to the signal.

The right panel of Fig.\, \ref{fig:chisquare} shows a correlation between the $\chi^2$ values obtained from the best fits to the ALMA images and those obtained from the SEDs. Such a correlation is expected because the best-fitting solution is determined by simultaneously minimising the combined $\chi^2$ of the two datasets.
Given the qualitative treatment of the errors conducted in this study, we inform the reader that the errors we provide should be taken as approximate estimates, rather than real errors attributed to the physical quantity. 

\subsection{Optimisation procedure} 

Our code requires as input information the dynamical mass for the star, the SED and a high resolution (FWHM$<0.15$\, arcsec) ALMA continuum image in bands 6 or 7 for each star. Given this information, the input parameters are the parallax ($\pi$), the dynamical mass ($M_\star$), the interstellar absorption in the $V$ band ($A_V$), the inclination ($i$) and position angle ($PA$) of the disc, collected from the literature as explained in the next section. The free parameters in the fit are the stellar effective temperature (\teff) and radius ($R_*$), the accretion rate (\mdot), the total mass (gas + dust) of the inner and outer portions of the disc ($M_{\rm inner}$) and ($M_{\rm outer}$), the inner and outer radii of the inner disc ($a_{\rm min, inner}$) and ($a_{\rm max, inner}$), and the maximum radius of the grains in the inner disc ($R_{\rm max,grain,inner}$). Given these parameters, the code computes both the SED and the images of the disc in the continuum at the appropriate wavelength, taking into account the inclination and position angle of the disc, convolved for the beam as read from the header of the file.  

\begin{table}[t]
\caption{Output parameters and the ranges for random extraction.}
\centering
\begin{tabular}{lcc}
\hline
\hline
Parameter            & Range      & Unit \\
\hline
\teff                & $\pm$100   & K   \\
$R_*$          & $\pm$10\%  &  \RSun   \\
$\log{\dot{M}}$      & $\pm$0.4   & M$_\odot\,\rm{yr}^{-1}$ \\
$M_{\rm inner}$     & $\pm$20\%  &   \Msun  \\
$M_{\rm outer}$      & $\pm$20\%  &   \Msun  \\
$a_{\rm min, inner}$ & $\pm$10\%  &  au   \\
$a_{\rm max, inner}$ & $\pm$10\%  &  au   \\
$\log{R_{\rm max,grain, inner}}$ & $\pm$0.6& cm  \\
\hline
\end{tabular}
\label{tab:range}
\end{table}

Optimisation was achieved by minimising the overall normalised $\chi^2$ value. The code allows for both manual optimisation (setting the number of random iterations to zero) and an arbitrary number of random iterations. They were obtained by considering random sets of values of the free parameters with uniform distributions within a given range from the original values that depend on the quantity considered (see Table \ref{tab:range}). Whenever the $\chi^2$ value obtained in this way was lower than that obtained with the original value, the best set of values is saved and used in the next iterations. The random walk allows progression towards a minimum solution. In practice, manual optimisation was used as a first step to provide the best centre of the images and a first approximation of the parameters. Subsequently, an automatic optimisation was launched until there was no further reduction of the value $\chi^2$ in the last 100 iterations. Finally, we again launched the automatic procedure by halving the ranges of variations used in the random extractions with respect to the values listed in Table \ref{tab:range}. In this case, we stopped the procedure only after there was no reduction of $\chi^2$ in the last 100 iterations. The procedure was completed by estimating uncertainties on the best fit parameters derived from all solutions whose $\chi^2$ value was higher than the best value by less than 10\%. The parameters obtained through our analysis are listed in Tables \ref{tab:output_data} and \ref{tab:derived_data}.

\subsection{Covariances}

\begin{figure}[t!]
    \centering
   \includegraphics[width=0.85\columnwidth]{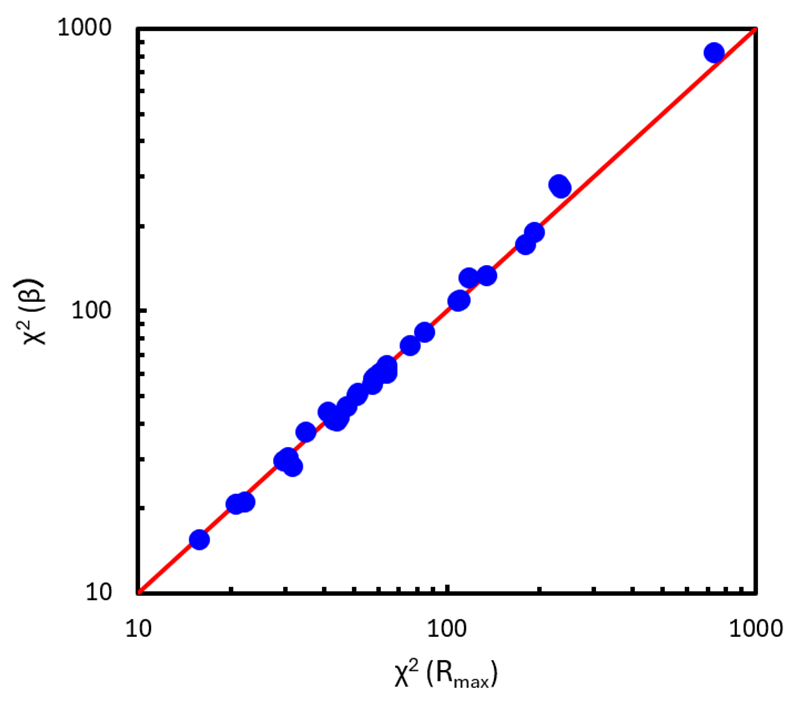}\\
   \includegraphics[width=0.85\columnwidth]{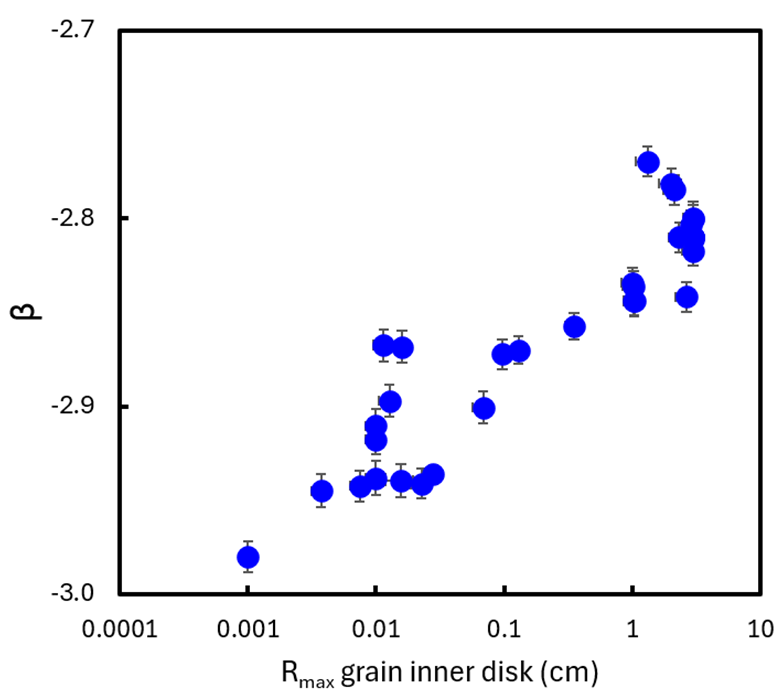}\\
   \includegraphics[width=0.85\columnwidth]{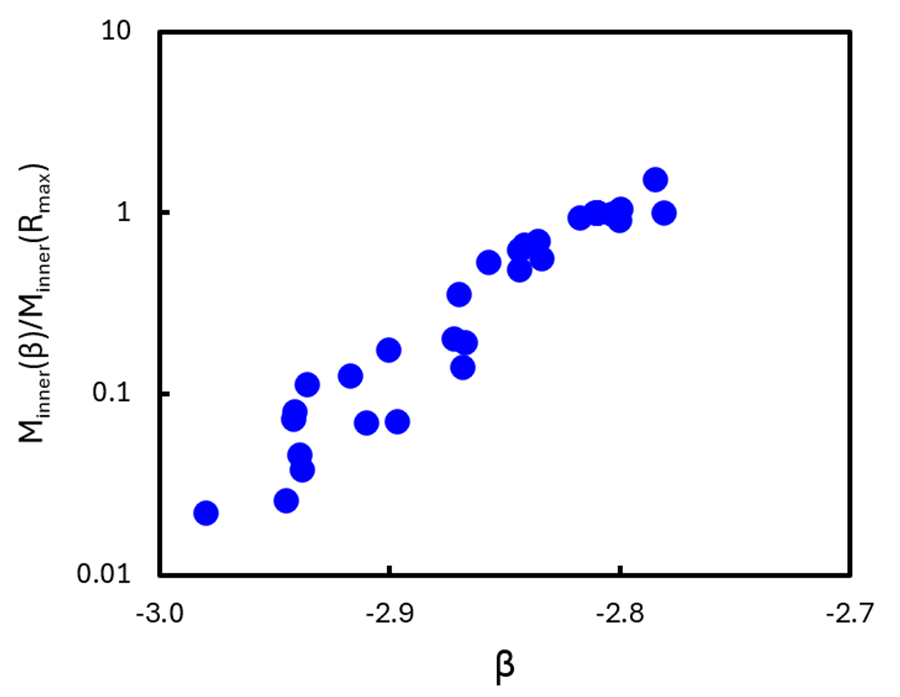}\\
    \caption{Upper panel: comparison between the $\chi^2$ for the best solutions obtained using the slope $\beta$ of the grain radius distribution for the inner disc and those obtained using a maximum grain size $R_{\rm max}$. Middle panel: slope $\beta$ as a function of $R_{\rm max}$ for the best solutions. Lower panel: ratio between the mass in the inner disc obtained using $\beta$ and $R_{\rm max}$ as a function of $\beta$.}
    \label{fig:slope}
\end{figure}

The covariance between the various quantities derived from our analysis can be examined using corner plots. We show three examples of these corner plots in Appendix~\href{https://doi.org/10.5281/zenodo.22792373}{C:Corner plots}. To derive them, we used 10,000 additional random extractions of the parameters around the best values from uniform distributions over the ranges given in Table \ref{tab:range}. We then evaluated the probability for each of these models as $p=\exp{( -\chi^2/2)}$ and normalised them to the peak value. We then plotted isocontours containing 68.3\% (1$\sigma$) and 95.5\% (2$\sigma$) of the probabilities in the posterior distributions.

In general, we expect that stellar radii and effective temperatures are anti-correlated; this is indeed found, though the range covered by the good solutions (probability $>0.01$) is actually small for both quantities. Also, we expect that the accretion rates are anti-correlated with effective temperatures and radii when we lack UV space data, which is however the case for only a few stars that are typically old and have low accretion rates.

More interesting in the present discussion is our choice to describe the different distribution of grain size in the inner and outer disc by considering the maximum value $R_{\rm max}$ (grain) for the grain size. An alternative choice could be to adopt a different slope $\beta$ of the grain distribution. We expect that this is a fully equivalent approach. To test the equivalence between these two approaches, we repeated the analysis for all discs in our sample by optimising $\beta$ rather than $R_{\rm max}$ (grain). As shown in Fig.\, \ref{fig:slope}, we found very similar best $\chi^2$ values using the two approaches. There is a clear correlation between $\beta$ and $R_{\rm max}$ (grain) for the best solutions. As expected, optimising $\beta$ rather than $R_{\rm max}$ results in different values of the inner disc mass. The ratio of the values obtained using $\beta$ and $R_{\rm max}$ is strongly correlated with the value of $\beta$. The other quantities (in particular the stellar ages) are not affected by this choice, although the best values may be slightly different due to our use of a Monte Carlo approach.

\subsection{Ages estimate}

Once the best solution was found, the ages were derived from a comparison of the radii with the isochrones of \citet{Baraffe2015} for the dynamical mass of the stars. In fact, during the pre-main sequence evolution, the radii of stars with masses similar to the Sun rapidly change, while effective temperatures stay roughly constant. The error bars given in Table \ref{tab:derived_data} are simply the internal error bars of our analysis.

Several studies found that the evolution of pre-main sequence stars is not very well reproduced by models that neglect stellar activity. Radii are actually less sensitive than effective temperatures to this issue; however, a comparison between the models by \citet{Baraffe2015} and those considering spots on the stellar surface (e.g., \citealt{Somers2015, Towner2025}) shows that the radii may be overestimated by as much as 7\%, and the ages underestimated by as much as 15\%, depending on the spot coverage. Although this introduces some uncertainty in the exact scale of ages used throughout this paper, it should not significantly affect the relative ranking.

\section{Results}
\label{sect:results}

\begin{figure*}[!t]
 \sidecaption
    \includegraphics[width=12cm]{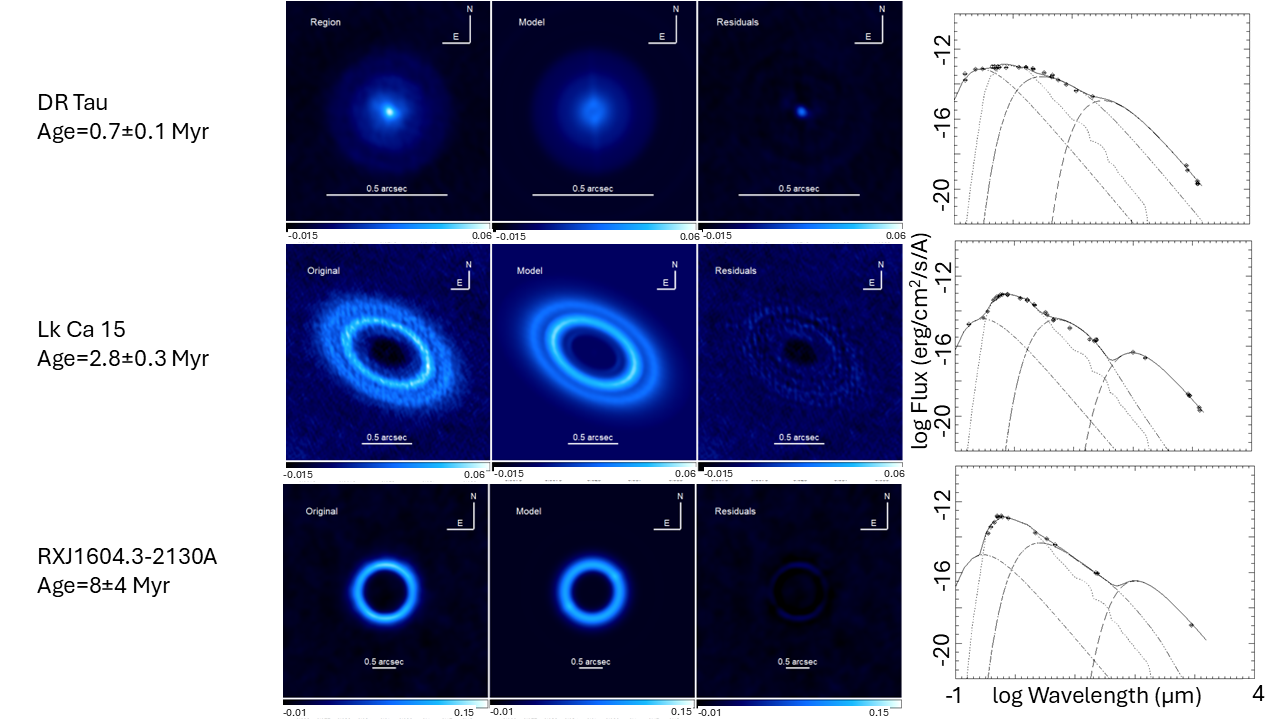}
    \caption{Results obtained with the model presented in this work for DR Tau (upper row), LkCa 15 (middle row), and RX\, J1604.3--2130A (lower row). From left to right: the original ALMA continuum image; the model of the ALMA image; the residuals after subtraction of the model from the original data, and the SED. In the last panel symbols with error bars are observations and the solid line is the fit. Individual contributions by the accretion region, the stellar photosphere, the inner disc, and the outer disc are shown as dot-dashed, dotted, long dashed and short dashed lines, respectively. }
    \label{fig:data_model_comparison}
\end{figure*}

Examples of the comparison between data and results of our model are shown in Fig.\, \ref{fig:data_model_comparison}, which shows the results obtained for a young (DR Tau), an intermediate (LkCa 15), and an old disc (RX\, J1604.3--2130A). Similar comparisons for the other stars are given in Appendix~\href{https://doi.org/10.5281/zenodo.22792373}{D:Fit to ALMA data and SED}. The surface density profiles are given in Appendix~\href{https://doi.org/10.5281/zenodo.22792373}{E:Disc surface density profile}. The evolution of the disc structure is quite evident in both the ALMA continuum image and the shape of the SED. The main changes are related to the progressive reduction of the contribution of the inner disc and of the accretion. We will return to this point in the next section. The best fit parameters obtained for all stars considered in this study are listed in Table \ref{tab:output_data}.

\subsection{Relations between observed quantities and derived parameters}

\begin{figure*}[!t]
    \centering
    \includegraphics[width=0.92\textwidth]{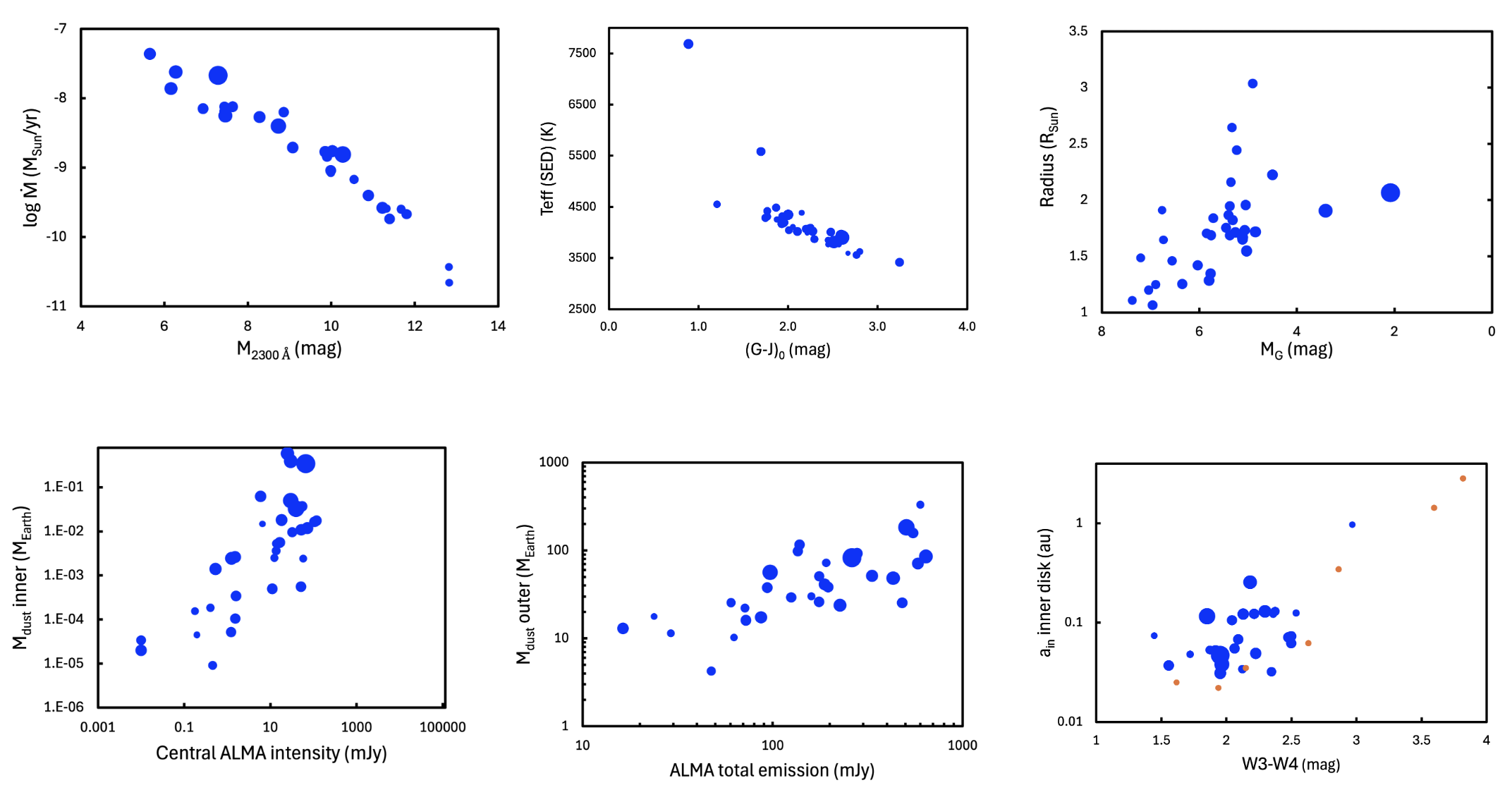}
    \caption{Comparisons between some input and output parameters used in our analysis. Symbols are proportional to the stellar radius. Upper left panel: absolute UV magnitude at 2300\, \AA\ and accretion rate $\dot{M}$; upper central panel: dereddened $G-J$ colour and effective temperature; upper right panel: absolute $G$ magnitude and radius (in this panel the diameter of the symbols is proportional to the effective temperature of the star); lower left panel: intensity of the ALMA image at the star position and dust mass of the inner disc; lower central panel: total flux measured by ALMA and dust mass in the outer disc; lower right panel: inner radius of the inner disc and $W3-W4$ colour according to WISE photometry (in this panel, open symbols are binaries).}
    \label{fig:param_comparison}
\end{figure*}

The parameters used in the proposed model depend mainly on some of the observational data, despite the fact that the final optimisation considers the total value of $\chi^2$. We show some of the correlations in Fig.\, \ref{fig:param_comparison}. 

There is a very tight correlation between the accretion rates and UV fluxes that we may represent as the absolute magnitude at 2300\, \AA, obtained from the observed flux corrected for the interstellar reddening and distance (see the upper left panel of Fig.\, \ref{fig:param_comparison}). The actual magnitude we considered is an approximation, where the value was obtained by interpolation of fluxes observed at other UV wavelengths.

The effective temperatures of the stars are constrained by the shape of the SED at optical wavelengths, which can be characterised, for example, by the $B_P-R_P$ colour; similar results are obtained using the $G-J$ colour. As expected, the correlation we obtain is narrow (central upper panel of Fig.\, \ref{fig:param_comparison}). Since the range of temperatures of most pre-main sequence stars in the range of mass considered in this study is not large, the stellar radius essentially depends on the absolute magnitude, as shown in the upper right panel of Fig.\,\ref{fig:param_comparison}, where the size of the symbols show the residual dependence on the effective temperature.

\begin{figure}[t!]
    \centering
    \includegraphics[width=\columnwidth]{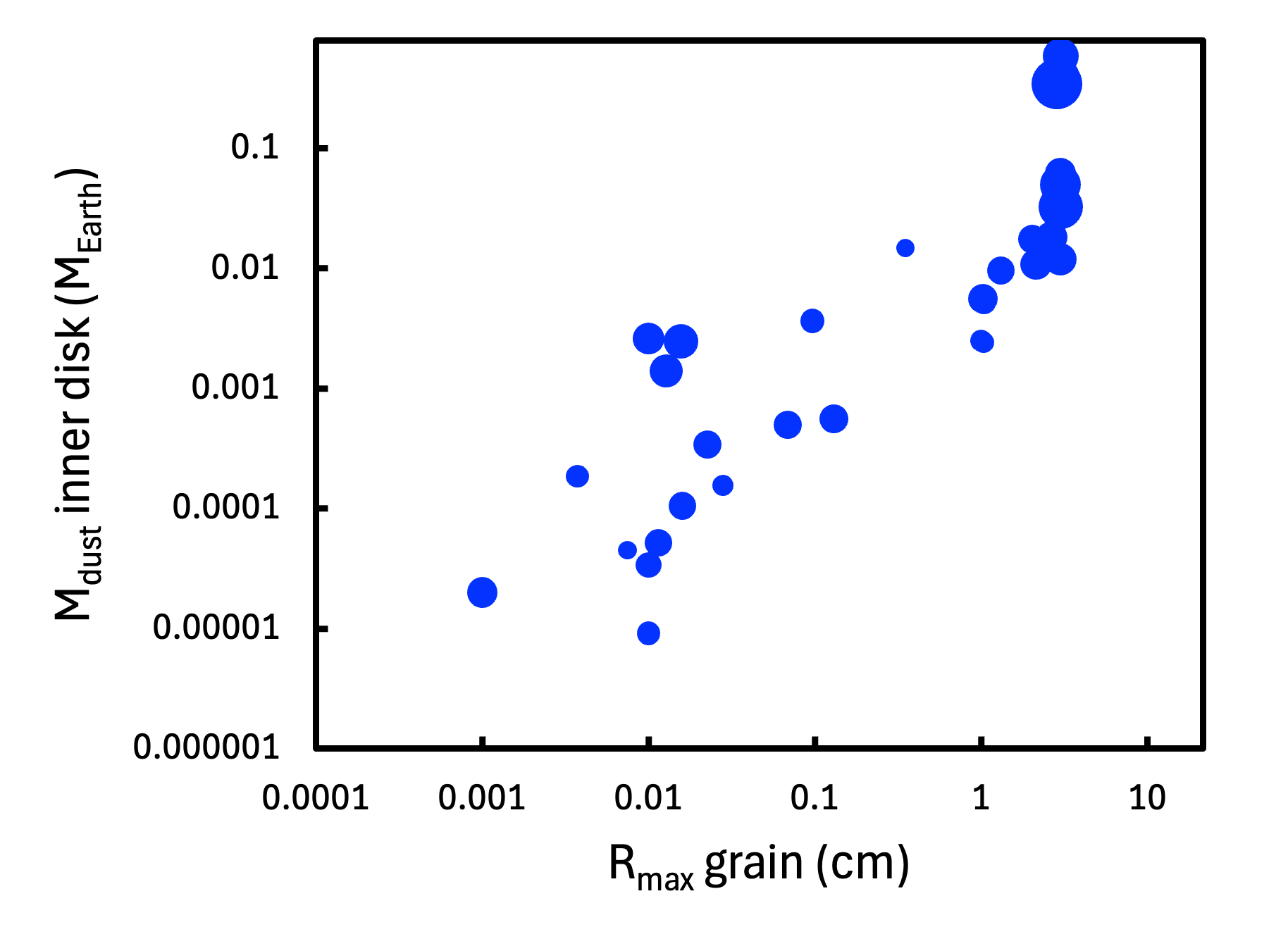}
    \caption{Correlation between the maximum radius for grains and dust mass in the inner disc. Symbols are proportional to the stellar radius.}
    \label{fig:rmax_grain}
\end{figure}

The dust masses of the inner and outer regions of the disc are strongly related to the continuum flux measured on the ALMA images. In particular, the inner-disc dust mass correlates with the flux measured at the centre of the image.
For sources with weak or undetected central ALMA emission, the inner-disc dust mass is instead constrained primarily by the SED. In these cases, the model adjusts the abundance of small dust grains to reproduce the near-infrared emission, thereby determining the inner-disc dust mass. However, a low value of the maximum grain size is required to simultaneously produce weak sub-mm emission from the inner disc. Since most of the dust mass is in large grains, the dust mass of the inner disc is mainly driven by the maximum radius that must be considered for the grains in the inner disc in order to reproduce the very low emission observed in the ALMA continuum images in the central region of some of the stars though simultaneously matching the SED in the mid infrared. In our models, this effect is only in part offset by the higher gas-to-dust ratio for the inner discs that we obtain in these cases (see Appendix \ref{Sect:equations}).
The central region contributes only marginally to the overall flux that is more closely correlated with the mass in the outer disc, which contains most of the dust (lower left and lower central panel of Fig.\, \ref{fig:param_comparison}). 

Finally, the inner radius of the inner disc determines the peak temperature of the dust and is then strongly correlated with thermal infrared colours. For example, a tight correlation is obtained with the $W3-W4$ WISE colour, that is, with the ratio of fluxes at 12 and 22 $\mu$m\footnote{\url{https://www.astro.ucla.edu/\, wright/WISE/passbands.html}}, as shown in the lower right panel of Fig.\, \ref{fig:param_comparison}.

In conclusion, the main properties of the stars and of the discs are strongly correlated with simple observables, allowing a simple physical interpretation and implying that the parameters we obtain are quite robust.

\subsection{Comparisons with literature values}

\begin{figure}[htb!]
    \centering
    \includegraphics[width=0.9\columnwidth]{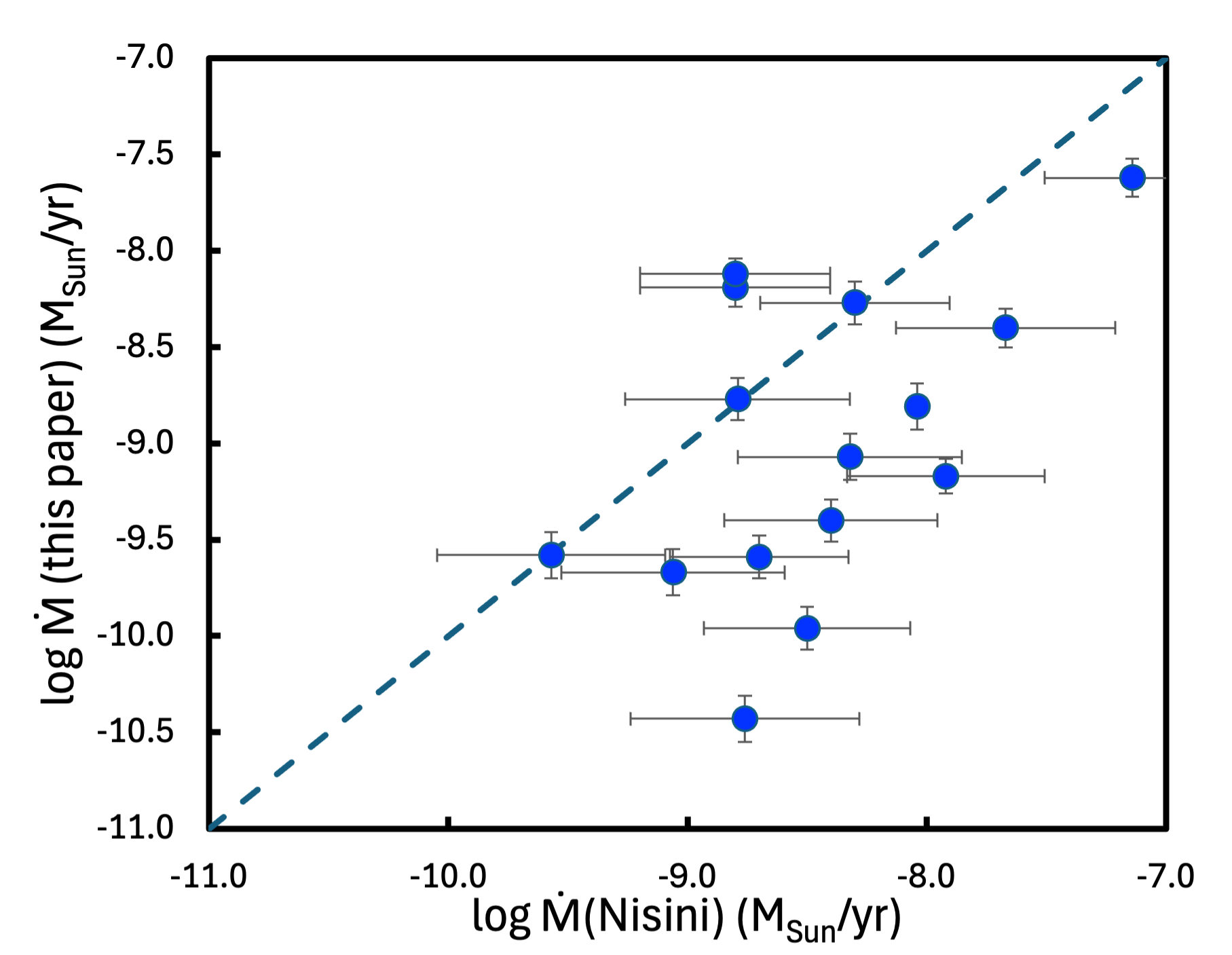}
    \caption{Comparison between the accretion rates obtained in this paper with those by \citet{Nisini2018}. The dashed line represents equality. }
    \label{fig:mdot_comparison}
\end{figure}

\begin{figure}[htb!]
    \centering
    \includegraphics[width=0.9\columnwidth]{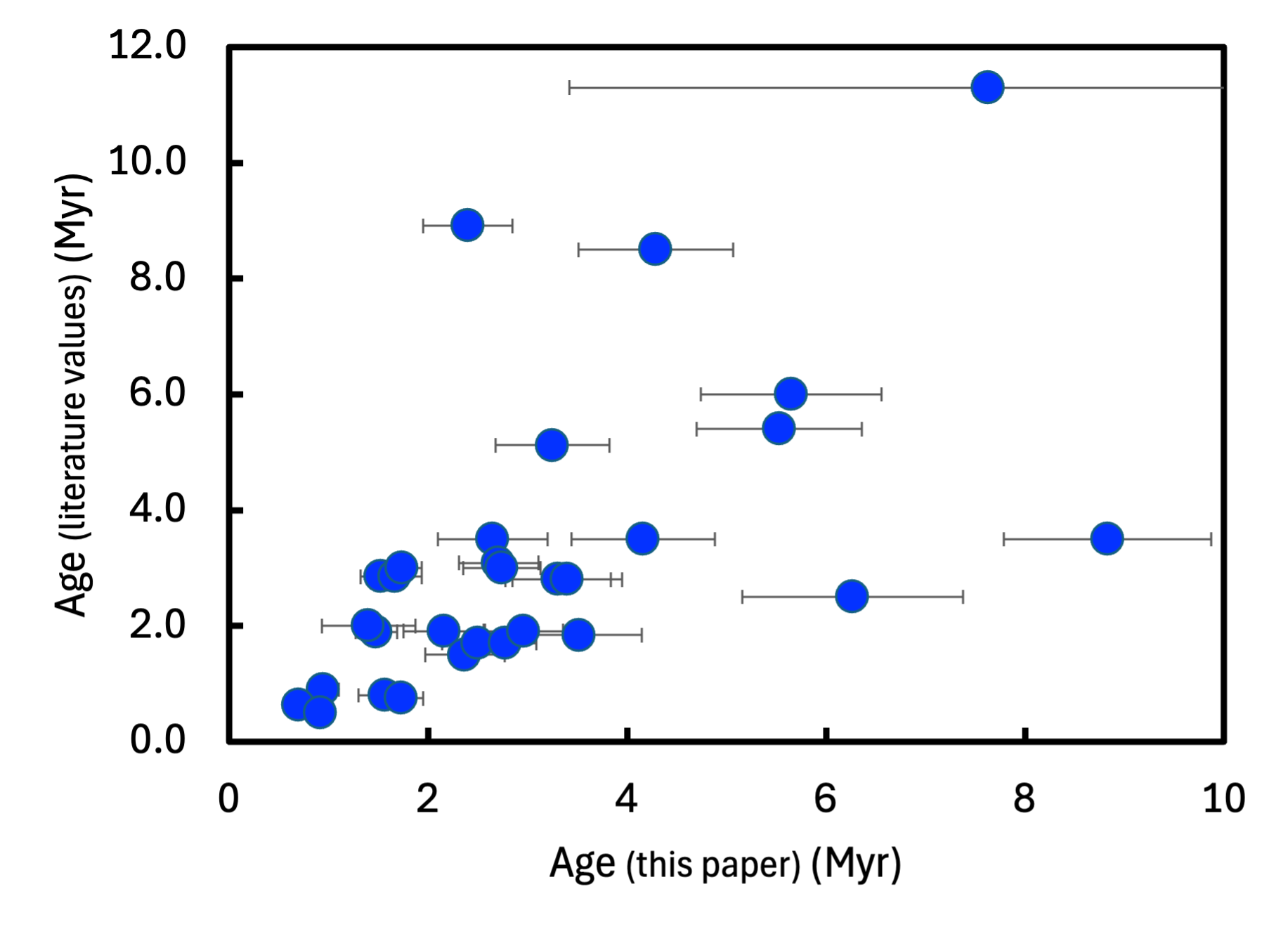}
    \caption{Ages derived in this paper compared with literature values.}
    \label{fig:age_comparison}
\end{figure}

In the following, we compared some of the quantities derived from our analysis with values in the literature. Effective temperatures can also be derived from spectra. We found offsets and individual differences that depend on the study we considered. We have eight stars in common with \citet{Flores2022}. They give temperatures derived from optical (from their search in the literature) and near infrared (their own study). Our values are higher by 98$\pm$82\, K (r.m.s. scatter of 231\, K) when using the optical spectra and 269$\pm$79\, K (r.m.s. scatter of 222\, K) when using the near infrared spectra. In contrast, our temperatures are systematically lower than those derived by other studies: by 99$\pm$39\, K (r.m.s. scatter of 123\, K) with \citet{Biazzo2017} (10 stars in common) and by 79$\pm$134\, K (r.m.s. scatter of 403\, K) with \citet{Gangi2022}. Finally, our temperatures are similar to those of \citet{Pittman2025}, with a mean difference of -11$\pm$64\, K, r.m.s. scatter of 202\, K, 10 stars in common. Although it is not obvious that our temperatures are better than those derived from spectroscopy, they are at least derived homogeneously for all stars in our sample. We noticed that adopting higher temperatures by 100\, K the radii would be reduced by 5\% and the ages would be higher by about 15\%.

We compared the measurements of the accretion rates determined in our study with the values of \citet{Nisini2018} mainly based on high-resolution spectroscopic data (mainly from X-Shooter data) (see Fig.\, \ref{fig:mdot_comparison}). We found a good correlation but a significant offset ($-0.69\pm 0.16$ dex with a standard deviation of 0.64\, dex, 15 stars in common). The agreement is also fairly good with the values estimated by \citet{Pittman2025}, with a mean difference of -0.62$\pm$0.19\, dex, r.m.s. scatter of 0.65\, dex over 10 stars in common. We also compared our values with those recently determined by \citet{Delfini2025}  from the {\it Gaia} XP spectra (prism low resolution), which are available for 27 of the stars considered here. In most cases, stars lacking the determination by \citet{Delfini2025} have old ages and low accretion rates. There is a fairly good correlation between the two sets of values that is well represented by a simple offset. Our estimates are lower by $0.29\pm 0.12$\, dex, with a standard deviation of 0.64\, dex. The standard deviation value is largely explained by errors in both determinations (mean quadratic values of 0.11 and 0.46\, dex in our and \citealt{Delfini2025}, yielding an expected scatter of 0.47 dex) with some contribution due to temporal variability of the accretion rates.  We noticed that while we determined the accretion rates directly from the accretion luminosities using space UV data, both \citet{Nisini2018} and \citet{Delfini2025} used the calibration of the equivalent width of the emission of $H\alpha$ of \citet{Alcala2017}.

Regarding the age estimates, Fig.\, \ref{fig:age_comparison} compares the ages derived in our study with the values from the literature (see Table \ref{tab:derived_data} for the references used for individual stars). For some stars, the adopted age corresponds to the estimated age of the stellar group of which the star is a member. The agreement is fairly good for most stars. We obtained younger ages for CQ Tau and PDS 66, and older ones for 2MASS J16120668--3010270 and V1094 Sco. The age given of CQ\, Tau by \citet{Vioque2018} (8.9\, Myr) is much older than the age of MWC 758 (3.5\, Myr) that belongs to the same kinematics group. We give almost the same age for the two stars. The age of 8.5\, Myr for PDS 66 given by \citet{Ribas2023} is older than ours ($4.3\pm 0.8$\, Myr). \citet{Asensio-Torres2021} gives an age of $3.1\pm 0.9$\, Myr for this star and various authors \citep{Torres2008, Murphy2013, Dickson-Vandervelde2021} found that it is a member of the $\epsilon$\, Chamaeleon association, which age is estimated at 3.8\, Myr. Our age of 2MASS J16120668–3010270 ($9\pm 1$\, Myr) is older than the value of 3.5 Myr given by \citet{Trapman2025}. The star belongs to the Upper Scorpius association and our age agrees well with recent estimates of the age of the association \citep{Pecaut2012, David2019, Towner2026}. Finally, we give an age of $6.3\pm 2.5$\, Myr for V1094\, Sco that is older than the value of \citet{vanTerwisga2018} (2.5 Myr). V1094 Sco is  located in the young Lupus Star Forming Region. Our estimate has quite large errors, and the star may be younger than we estimated.

Detailed comparisons for quantities related to discs are more difficult due to the large number of parameters involved, the arbitrary approximations we used in our modelling, and the different definitions used for inner and outer discs. For these reasons, the comparison is mostly qualitative. A low value of  central emission in ALMA images indicates the presence of a cavity, which is a common feature observed in transitional discs \citep{vanderMarel2023}. Using our model, this is represented by a low value of the maximum grain radius or by a steeper slope in the distribution of grains with radius in the inner disc (see Fig. \ref{fig:param_comparison}). We found a correlation between the disc classification of \citet{Parker2022} (available for 14 stars) and the values of $R_{\rm max}$-grain or the slope in the grain distribution in the inner disc: low values of these last quantities are associated with rim discs and high values with ring/spiral discs, although there are a couple of exceptions because we obtain low values of $R_{\rm max}$ for the inner discs of GM Aur and Sz 129 that are classified as ring/spiral discs.

\begin{figure}[htb!]
    \centering
    \includegraphics[width=0.9\columnwidth]{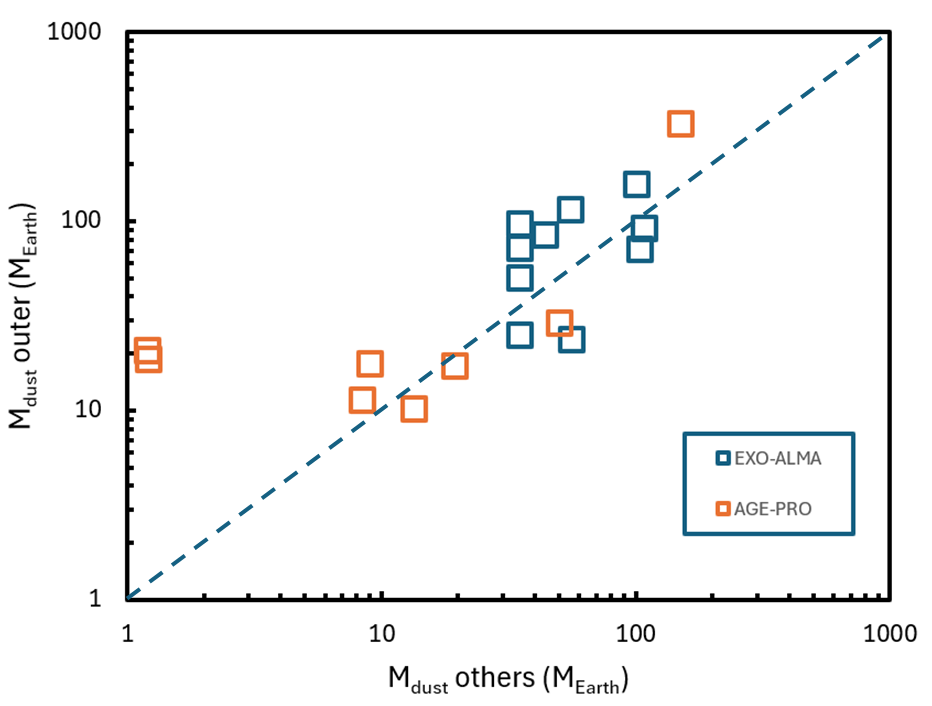}\\
    \caption{Comparison between the outer disk dust masses (the outer disk mass is used as a proxy of the total mass) obtained in this paper and those from the {\tt EXO-ALMA} (\citealt{Curone2025}: open blue circles) and {\tt AGE-PRO} (\citealt{Agurto-Gangas2025, Deng2025}: open orange squares) programmes. The dashed line represents equality. }
    \label{fig:comp_dust_masses}
\end{figure}

The total mass in dust in the discs (almost all in the outer disc) is directly related to the total flux in the ALMA continuum images, with a weak dependence on disc temperature and grain opacity. We compared our values with those obtained in the {\tt EXO-ALMA} \citep{Curone2025} and {\tt AGE-PRO} \citep{Agurto-Gangas2025, Deng2025} programmes (see Fig. \ref{fig:comp_dust_masses}). There are ten stars in common with {\tt EXO-ALMA} and eight with {\tt AGE-PRO}. We found that on average the masses determined in this paper are larger by $0.08\pm 0.10$ dex with an r.m.s. scatter of 0.27 dex with respect to {\tt EXO-ALMA} and $0.20\pm 0.35$ dex with an r.m.s. scatter of 0.57\, dex with {\tt AGE-PRO}. Most of the scatter in the last comparison is due to the two discs with lowest dust mass (Sz 77 and 2MASS J16202863--2442087), for which we derived higher dust masses. Both cases have compact and barely resolved discs with noisy density profiles in the outer regions. Our results are then sensitive to the exact zero point of the ALMA images.

\begin{figure}[htb!]
    \centering
    \includegraphics[width=0.9\columnwidth]{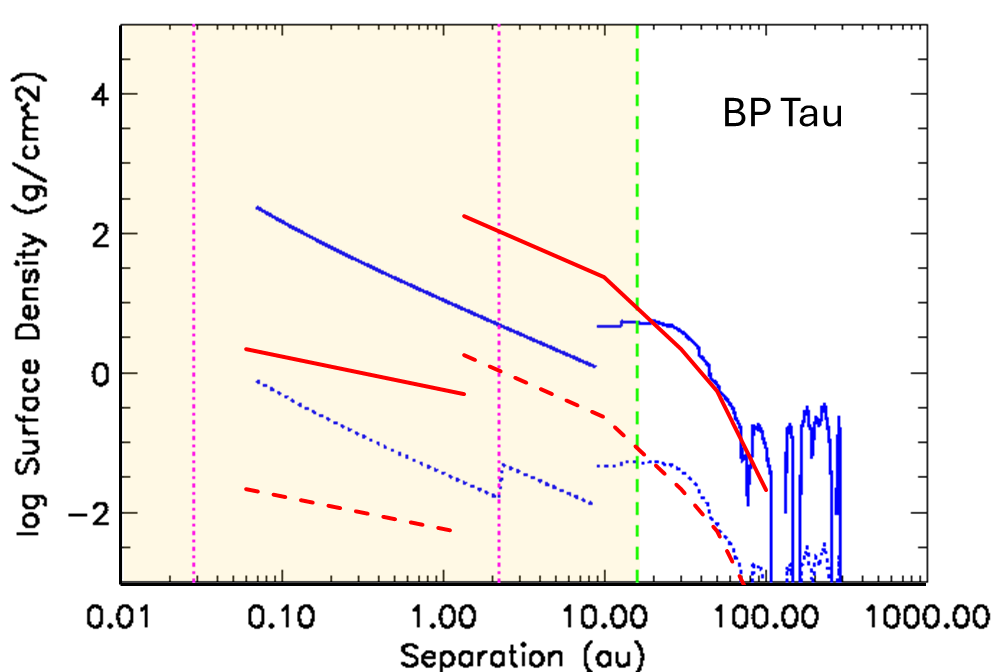}\\
    \includegraphics[width=0.9\columnwidth]{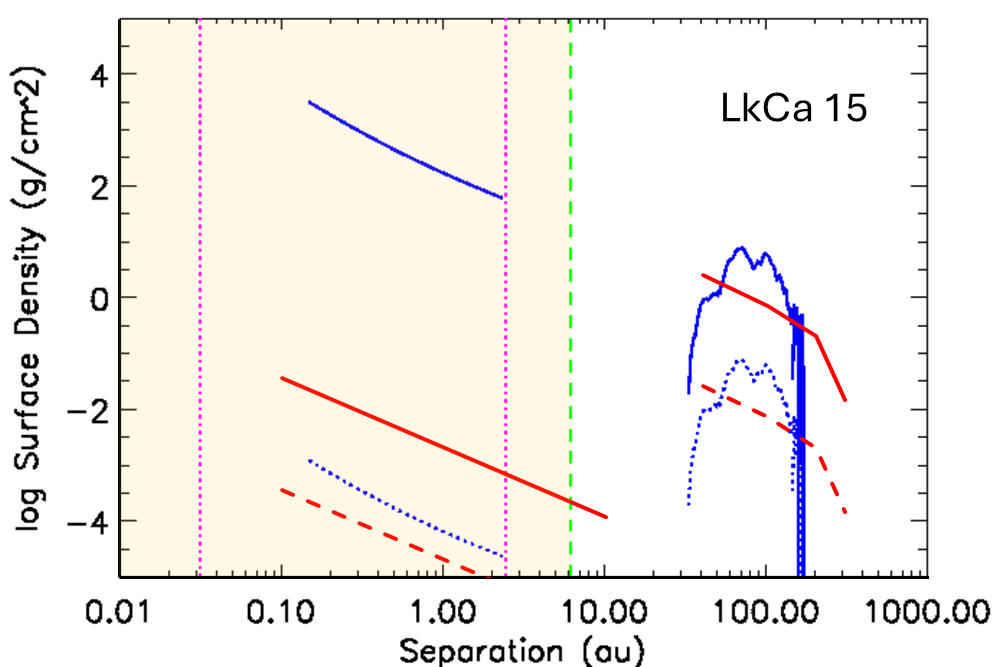}\\
    \includegraphics[width=0.9\columnwidth]{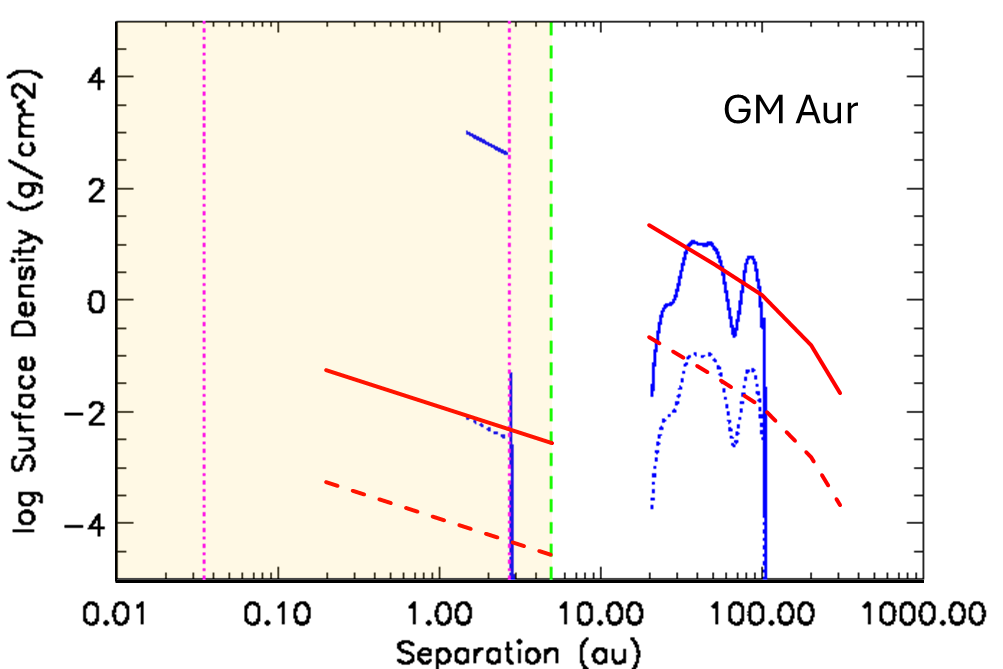}
    \caption{Radial surface density profiles from our model (blue lines) and from\citet{Woitke2019} (red lines) for BP Tau (upper panel), LkCa 15 (middle panel), and GM Aur (lower panel). Solid lines are the gas surface density; dashed lines are the surface density in grains. Vertical magenta dotted lines mark the location of the sublimation distances for silicates and ice. The vertical green dashed line represents the resolution of the ALMA images and the shaded area is the region unresolved in these images. }
    \label{fig:woitke}
\end{figure}

Finally, there are three stars (BP Tau, LkCa 15, and GM Aur) in common with the paper of \citet{Woitke2019} who made an analysis of several discs considering both SED and sub-millimetre observations available at that epoch and a radiative transfer approach. Several assumptions are different between \citet{Woitke2019} and our analysis. Anyway, we attempted a comparison. The surface density profiles obtained in the two studies are shown in Fig. \ref{fig:woitke}. We highlight here that given the considered observables, in both models the measured quantity is the dust density, although in both cases the disc surface densities (including both gas and dust) are also given. To estimate this last, \citet{Woitke2019} assumed a constant gas-to-dust ratio of 100 along all the disc. In our model, we considered a constant gas-to-dust ratio of 100 for the outer disc, while for the inner disc it depends on the distance from the star (only silicate grains are considered between the silicate and ice sublimation distances) and on the actual distribution of grains with size (as expressed by the parameter $R_{\rm max}$; see Appendix \ref{Sect:equations}). The adopted method implies that the gas-to-dust ratio in the inner disc may be much higher than 100. As shown in Fig.~\ref{fig:woitke}, there is a good agreement for the outer discs, for which both studies assumed a gas-to-dust ratio of 100. Indeed, in both models, most of the dust mass is in large grains and the emissivity in the millimetre bands is roughly proportional to their mass. The (small) differences found are due to the availability of higher resolution ALMA images that allowed us a higher resolution in the radial profiles. For the inner disc, the dust densities derived in this paper are higher by about an order of magnitude, with significant star-to-star variations. The differences found for the inner discs depend in part on the assumptions on their radial extension, which we found to be lower than that considered by \citet{Woitke2019}. Once this is considered, the logarithms of the dust masses in the inner disc according to our model are higher by $0.67\pm 0.11$ dex (r.m.s. of 0.19 dex). In addition, since the inner disc is optically thick at short wavelengths, the estimate of the surface density is uncertain because we often are in a saturation regime. In principle, a radiative transfer approach such as that considered by \citet{Woitke2019} should be more appropriate than our simplified one. However, we think that the possibility to rapidly estimate properties of many discs using the method considered in this paper may offset this disadvantage in the analysis of rather extensive statistical samples such as that considered here.

\section{Discussion}
\label{sect:discussion}

\begin{figure*}[htb!]
    \centering
    \includegraphics[width=\textwidth]{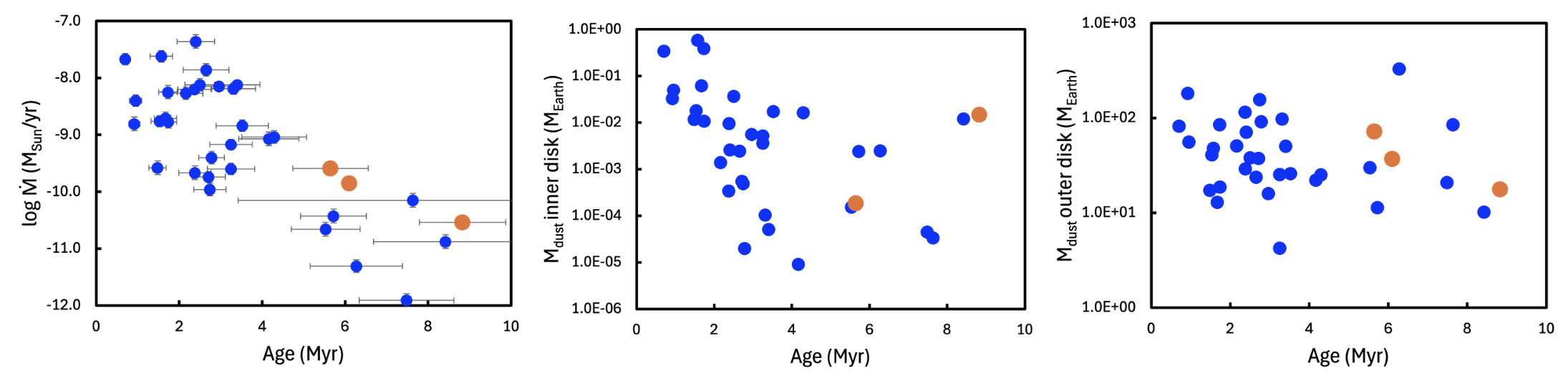}
    \caption{Trends with age for the mass accretion rate (left panel), dust mass of the inner disc (middle panel), and dust mass of the outer disc (right panel). In all panels, filled orange circles are stars hosting accreting planets, blue circles are other stars. }
    \label{fig:age_dependence}
\end{figure*}

\begin{figure}[htb!]
    \centering
    \includegraphics[width=0.9\columnwidth]{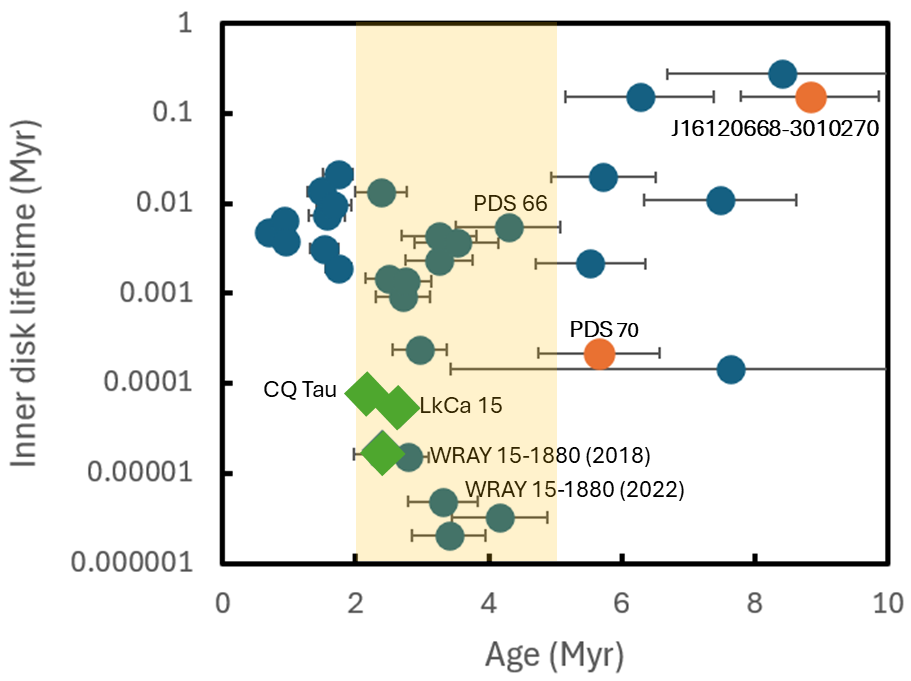}\\
    \includegraphics[width=0.9\columnwidth]{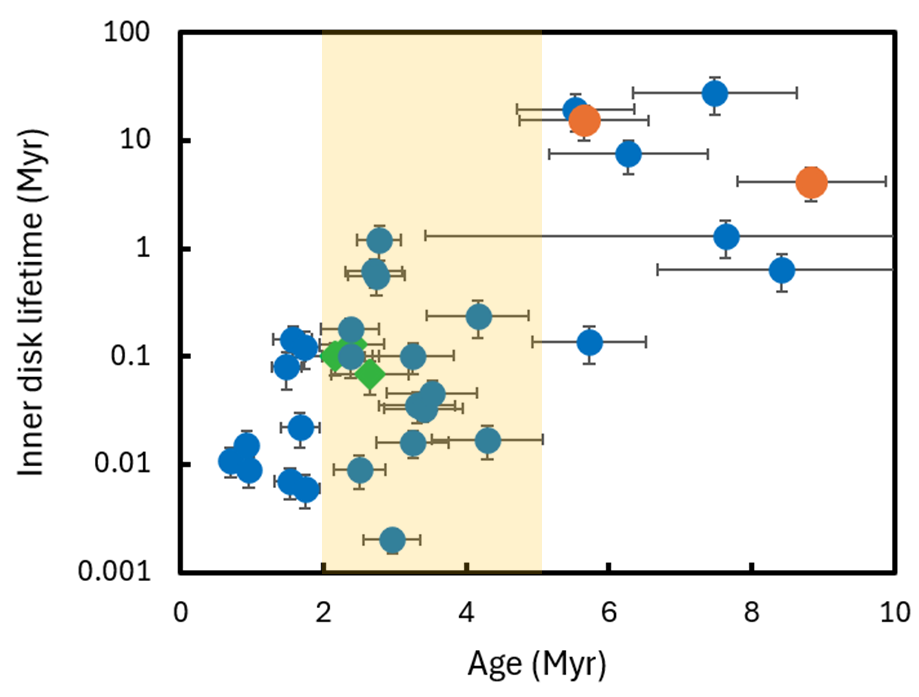}\\
    \caption { Lifetime of the inner discs as a function of age assuming a constant gas-to-dust ratio of 100 ({\it top}) and using the inner disc masses obtained by our model ({\it bottom}). In both panels filled orange circles are stars hosting accreting planets, green diamonds are discs with large asymmetries, blue circles are other stars. A few stars suspected to host planets are labelled in the first panel. The shaded area refers to the epoch between 2--5\, Myr where giant planet cores are likely building-up. }
    \label{fig:disc_lifetime}
\end{figure}

\subsection{Trends with age}

Figure \ref{fig:age_dependence} shows the trends of the age estimates with the mass accretion rate, the dust mass of the inner disc and the dust mass of the outer disc. The observed trends are quite obvious in the first two panels and confirm some known observations \citep{Hartmann1998, Hernandez2007, Fedele2010}. In the last case, we should also notice that the stars considered in this study are biassed towards those that have prominent discs \citep{vanderMarel2023}. Although a large fraction of very young stars have prominent discs, less than a third of those older than 5\, Myr hosts a clearly detectable disc \citep{Rigliaco2025}. Once this is considered, some trend of the dust mass of the outer disc with age is also present, although less clear than for the inner disc. 

In general, these data suggest that this sample is dominated by young or transitional discs. Classical transitional discs are defined as protoplanetary discs with little or no near-IR excess ($\lambda<10$\, $\mu$m) and a significant excess comparable to the median of Taurus \citep{Furlan2006} at longer wavelengths \citep{Strom1989, Calvet2002, Calvet2005}. However, it must be noted that while there is some trend between the ratio of fluxes at short and long wavelengths in the IR with age, the scatter is very large. The large scatter may be attributed to the weak correlation between the inner radius of the inner disc and age within our sample.

\subsection{Inner disc evolution and the formation of giant planets}

Assuming a constant gas-to-dust ratio (e.g. at a value of 100 as in our model), we may divide the mass of the inner disc (here defined as the region within the ice line) by the accretion rate onto the star. We call this quantity disc lifetime. Insofar this approximation were correct, we expect that the lifetime of the disc is related to basic properties of a disc \citep{Lynden-Bell1974}. In fact, the viscous timescale for a thin disc is:
\begin{equation}
t_{\rm vis} \sim t_{\rm therm} (a/h)^2
\end{equation}
where $t_{\rm therm}$ is the thermal timescale, $a$ the radius and $h$ the height of the disc. In turn, the thermal timescale is:
\begin{equation}
t_{\rm therm} \sim t_{\rm dyn} / \alpha
\end{equation}
where $t_{\rm dyn}$ is the dynamical timescale and $\alpha$ is the viscosity coefficient. Assuming $(h/a)=0.1$ and $\alpha=0.001$, we have:
\begin{equation}
t_{\rm vis} \sim 10^5 t_{\rm dyn},
\end{equation}
where $t_{\rm dyn}$ is roughly the orbital period. For a typical inner disc ($a\sim 0.3$\, au), $t_{\rm dyn}\sim 0.1$\, yr, and then $t_{\rm vis} \sim 10^4$\, yr (see also \citealt{Lodato2017, Alexander2023}).  

We indeed find that for ages $<2$\,Myr the lifetime of the inner disc is of the same order as the viscous timescale, as expected. Interestingly, the masses of the discs in the inner disc for the youngest stars of our sample align well around a single disc isochrone (as defined by \citealt{Lodato2017}) corresponding to a disc lifetime of about $10^4$\, yr (see Fig.\, \ref{fig:disc_lifetime}). 

However, the lifetime of the inner disc obtained this way is $<1000$ yr for 9 out of 15 of the discs around stars in the age range 2--5\, Myr; the values are so low as to imply some drawback in this interpretation. We noticed that our model does not really constrain the mass of the inner disc but only the mass in dust, and we infer the mass of the inner disc by assuming a given gas-to-dust ratio. However, in our models, we reconciled the low central emission in the ALMA images with the near infrared excess given by the SED assuming that these inner discs contain few large grains and effectively have large gas-to-dust ratios. The lifetimes of the inner discs in this age range estimated using the total masses we obtained with our model are much longer than assuming a constant gas-to-dust ratio for all discs, and this potential concern disappears as shown by a comparison of the two panels of Fig. \ref{fig:disc_lifetime}. In addition, we remind them that these discs might contain a significant mass in planetariums that do not contribute to the emission. 

Lifetime is instead long for some of the inner discs with ages $>5$\, Myr. However, we should also notice that in the oldest stars in our sample, a significant fraction of the UV excess may be due to chromospheric emission rather than to the accretion region \citep{Manara2013}. In this case, the accretion rates may be lower (and lifetime higher) than what is given by our combined star/disc model. Conversely, there is no clear correlation when we consider the total mass of the disc, which is very close to the mass of the outer disc.

\subsection{Stars hosting accreting planets}

In all panels of Figs.\, \ref{fig:age_dependence} and \ref{fig:disc_lifetime} we use different symbols for stars known to host accreting planets (PDS 70: \citealt{Keppler2018}; 2MASS J16120668--3010270: \citealt{Close2025}; and TYC 5709-354-1: \citealt{Close2025}). They have similar ages of approximately 6--8,Myr, corresponding to a stage at which both the stellar accretion rates and the inner-disc dust masses are substantially lower than those observed in the younger stars in our sample. So far, no planet has been firmly detected around stars with masses similar to the Sun and younger than 5\,Myr, although some candidates have been proposed as mentioned in Sect. \ref{sect:sample}. 

Both PDS 70 and 2MASS J16120668--3010270 exhibit cavities in their ALMA continuum emission that may be interpreted as evidence for a depletion of large grains in the inner disc. Similar structures are observed in several other discs in our sample, most of which are younger. However, a few stars older than 5\,Myr also show comparable features, with RXJ1852.3-3700 and RXJ1604.3-2130A providing the clearest examples. Both exhibit ring-like structures in their ALMA emission that are similar to those observed in PDS 70.
The first one has been observed with the SPHERE high contrast imaging but only in DPI mode. RXJ1604.3-2130A was studied by \citet{Canovas2017} without detection at a mass sensitivity of $\sim$2--3\,\Mjup from 22 to 115\, au. We noticed that the planets around PDS-70 and 2MASS J1612 are giant planets at large separations: PDS70b and c are at 21 and 34\, au, 2MASS J1612b is at 24\, au. Most giant planets are thought to be at a shorter separation and it is well possible that other stars host undetected giant planets.

\subsection{Discs with large residuals}
\label{sect:strong_asymmetry}

\begin{figure*}[t!]
\sidecaption
    \includegraphics[width=12cm]{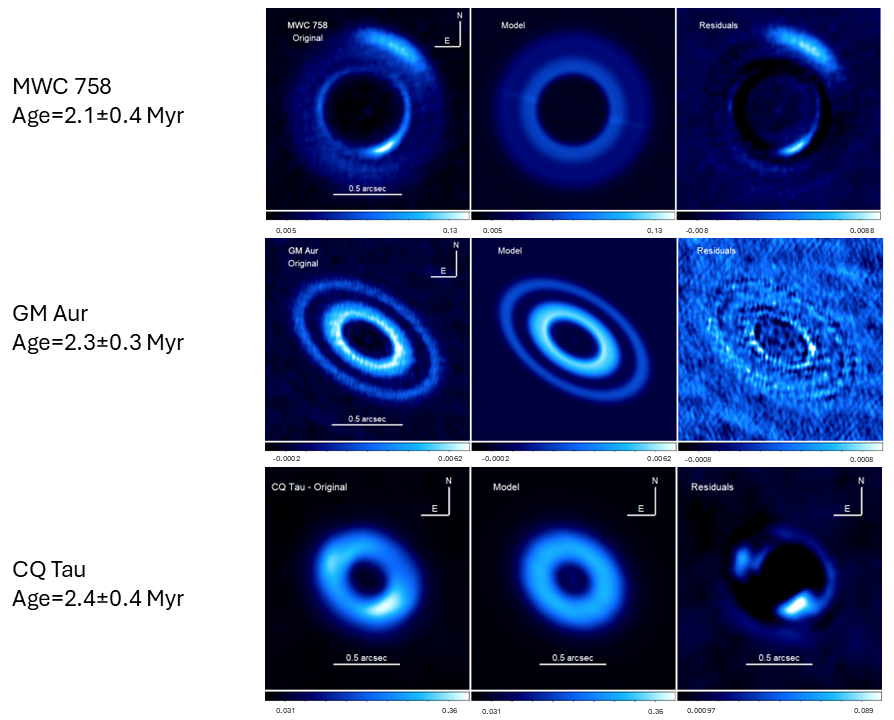}
    \caption{ALMA dust images of discs exhibiting strong residuals. From top to bottom: MWC 758, GM Aur, and CQ Tau. In each row, the left panel is the original ALMA image, the central panel is the model; and the right panel shows residuals. }
    \label{fig:strong_residuals}
\end{figure*}

MWC 758 and CQ Tau are surrounded by discs showing strong deviations from radial symmetry (see Fig.\ref{fig:strong_residuals}). The asymmetries were already noted (MWC 758: \citealt{Boehler2018, Casassus2019}; CQ Tau: \citealt{Ubeira2019}) and have been attributed to the presence of vortices that trap large dust grains, possibly related to the presence of unseen companions \citep{Boehler2018, Ubeira2019}. In addition, large residuals are also found for GM Aur. They are quite massive stars in the sample, with mass $>1.0$\, \Msun. Strong asymmetries are frequently observed in the discs around Herbig stars \citep{Kraus2009, Kluska2020, Varga2021, Booth2024} that are typically more massive than the solar type stars considered in this study. The results for MWC 758, CQ Tau, and GM Aur might fit within a picture where the discs around massive stars are more prone to develop strong asymmetries. We also noticed that our analysis underestimates the total emission and then the total mass of dust (in the outer disc) for these asymmetric discs.

\subsection{Relation to planet formation}

\begin{figure}[htb!]
    \centering
    \includegraphics[width=0.45\textwidth]{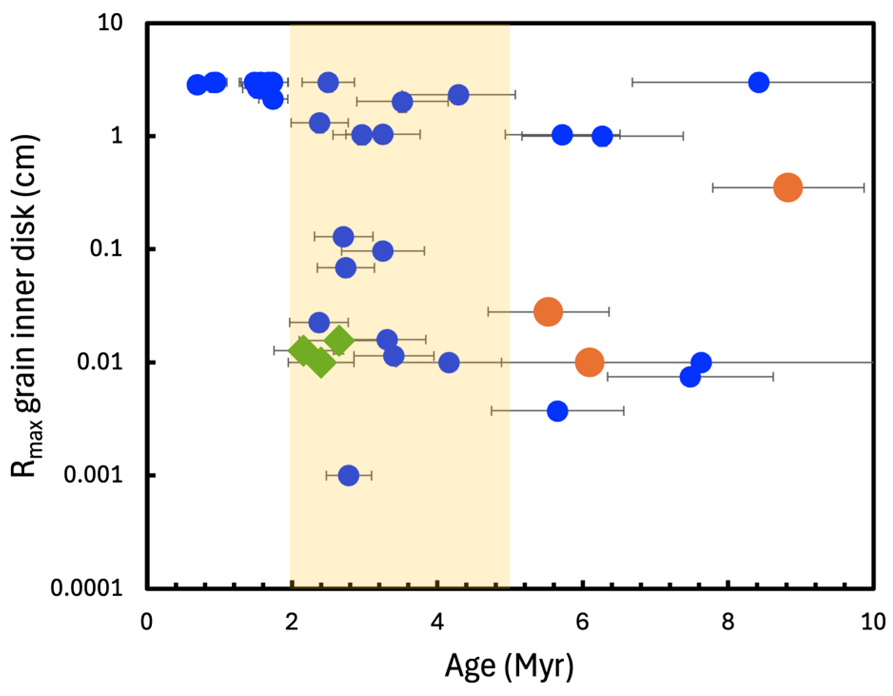}
    \caption {Maximum radius of grains in the inner disc as function of age. Filled orange circles are stars hosting accreting planets, green diamonds are discs with large asymmetries, blue circles are other stars. The shaded area refers to the epoch between 2--5\, Myr where giant planet cores are likely building-up.}
    \label{fig:disc_maxgrains}
\end{figure}

Figure \ref{fig:disc_maxgrains} shows the maximum radius of grains in the inner disc as a function of age. The figure highlights a relative lack of large grains in the inner parts of some discs around stars with ages in the range 2--5\, Myr, including the cases of stars with known accreting planets, with respect to discs around younger stars. We may quantify this depletion considering that the slope of the grain distribution ranges from about $-$2.8 to values lower than $-$2.95. Since we consider grains with radii from 0.1\, $\mu$m up to 3 cm, this implies that grains with larger radii are depleted by a factor of about 6 with respect to those of smaller radii. The depletion of large grains might be attributed to the presence of obstacles to their inward motion \citep{Zhang2018}, although other interpretations are also possible (e.g. \citealt{Zhang2015}). We also noticed that the strongly asymmetric discs present in the sample (MWC 758, CQ Tau, and GM Aur), yielding large residuals to the best fit ALMA continuum images, all have ages close to 2.4\, Myr, close to the transition between long and short lifetimes of the inner disc. The ages of these stars actually have uncertainties that are not negligible, so their almost identical ages might be a coincidence. If we neglect the presence of a companion to CQ Tau indicated by astrometry and adopt a mass of 1.4\, \Msun for the primary as indicated by disc kinematics \citep{Izquierdo2025}, the age would be 3.6\, Myr. However, these asymmetries might be explained as a result of dust traps  \citep{Pinilla2012, vanderMarel2013, Owen2017} and may represent a response of discs to the formation of large planetary cores. If we interpret the lack of large grains as evidence for the presence of giant planets even when this is not directly detected \citep{Zhang2018}, this would suggest that the minimum time required for the giant planet formation in the solar type stars considered in our study (which is biassed towards long-living discs) is $\sim 2$\, Myr. Finally, four of the nine stars older than 5\, Myr also have large grains in the inner disc, despite the low dust content of the inner disc ($<0.01$\, M$_{\rm Earth}$).

To further discuss this issue, we noticed that dust particles make up a small fraction of the total disc mass (nominally 1\%) and thus most likely follow the typically subKeplerian orbits of the gas. Lacking radial pressure support of the gas, they slowly fall onto the central star in a process called radial drift \citep{Weidenschilling1977}. By acting as a barrier, the core/planet may effectively filter incoming dust from the outer disc, altering what reaches the inner disc \citep{Kalyaan2023}. Although large grains can be efficiently trapped at the outer edges of carved spaces by planets, smaller particles tend to closely follow the motion of the gas and diffuse through the trap — a process called dust filtration \citep{Paardekooper2006, Rice2006}. \citet{Weber2018} showed that only particles above a certain Stokes number (or aerodynamic size) can be effectively stopped from moving inward for a given diffusivity. From their calculations, we derived that the order of magnitude for an appreciable effect is $r_{\rm grain}\geq 0.03$ cm and it is related to a core/planet to star mass ratio of $q\approx 5\times 10^{-4}$, which is about $\approx 0.5$\, \MJup for a solar type star. The threshold in the grain radius for efficient dust filtration compares well with that emerging from Fig. \ref{fig:rmax_grain}, but we warn the reader that model estimates for both this threshold and the mass of the core/planet originating the gap are uncertain, depending on the adopted viscosity of the disc and on how particle diffusion is considered. Tracking the size distribution of the dust throughout the evolution of a protoplanetary disc is thus key in quantifying the dust filtering efficiency of gaps in the disk gas distribution opened by planets. On the other hand, dust grains undergo continuous collisions, leading to fragmentation (break-up) and coagulation (sticking) of the interacting particles, depending on the material properties and collision velocities \citep{Birnstiel2024}. Following these phenomena is then important to establish the ability of a gap in the gas distribution due to a planet to affect the size distribution of grains in the inner disc.

Given the uncertainties in the underlying physics, the minimum planetary mass required for efficient dust filtration remains poorly constrained. For example, \citet{Perez2019} considered a mini-Neptune with a mass of 12\,\MEarth ($\approx 0.04$\, \MJup) sufficient to open a gap in the gas distribution of a protoplanetary disc. However, it remains unclear whether a planet of this mass would produce observable differences in the dust-grain distribution between the inner and outer disc regions. A mass of 12\,\MEarth is comparable to the Jupiter core prior to gas accretion in the model of \citet{Pollack1996}. Given the uncertainties in current planet-formation models, however, a broader mass range of 10--30\,\MEarth may be more appropriate.
Updated hydrodynamical simulations and a discussion of "dust filtering" due to the gap in a disc generated by a planet can be found in \citet{Pfeil2025}. These authors considered 2D multifluid hydrodynamic simulations of planet-disc systems with dust coagulation and fragmentation. They found that indeed results are sensitive to the grain-grain physics and then to the properties of the disc (e.g. diffusivity). They concluded that a planet with a mass of 0.6\, \MJup likely produces a significant depletion of large grains in the inner disc while the "dust filtering" is likely not enough to compensate for coagulation (which prevents the shortage of large grains in the inner disc) for planets as small as 0.1\, \MJup (the expected size of the core of Jupiter after 1-2\, Myr). The  shortage of large grains in the inner discs indicates an efficient "dust filtering" in more than half of the discs in our sample with ages $>2$ Myr: this would then suggest the presence of a giant planet (or a core) more massive than about 30\, \MEarth in this fraction of the discs in this age range in our sample.

In this context, it may be interesting to compare our results with the estimated properties of embedded planets given by \citet{Ruzza2026}. They considered discs observed within the exoALMA survey \citep{Teague2025}; ten of them are included in our study. They selected eight discs with indications for embedded planets and used the {\tt DB-Nets2.0} software to deduce their properties, including their mass. In this sample, we found that \citet{Ruzza2026} proposed planets with masses $>2$ \Mjup for all discs for which we obtained values of $R_{\rm max}<0.03$ cm in the inner disc. However, they proposed only a very small planet (mass of 0.29\, \MJup) for RX J1615.3-3255 for which we obtained $R_{\rm max}=0.069\pm 0.012$ cm. We obtained a high value of $R_{\rm max}=3$\, cm for PDS 66 that was not considered by \citet{Ruzza2026} because it lacked any indication of the possible presence of a planet. 

Finally, for WRAY 15-1880 (for which we obtained a low value of $R_{\rm max}=0.013\pm 0.002$\, cm for the inner disc) they propose the presence of a planet as massive as 0.3\, \MJup in the disc gap. Their estimate is consistent with the candidate companion proposed by \citet{Rigliaco2026} using high contrast imaging, with a mass in the range 0.3--7.6\, \MJup at 25\, au. 

In summary, the analysis suggests that large grains are depleted in the inner regions of several discs older than $\sim 2$\, Myr, which potentially indicates dust filtration by forming giant planets and suggests that planet formation may begin within $\sim 2$\, Myr. Planet-induced gaps can efficiently filter large grains while smaller particles remain coupled to the gas, with the efficiency strongly depending on the mass of the planet, the diffusivity of the disc, and the coagulation/fragmentation processes of the grains. The observed depletion in more than half of the discs older than 2\, Myr may therefore point to the presence of giant planets or planetary cores more massive than $\sim 30$\, \MEarth, although the exact mass threshold remains uncertain.

As a final consistency check, we observed that 9 of the 15 stars in our sample with ages in the range 2--5,Myr show a shortage of large grains in the inner disc, corresponding to a fraction of $0.60\pm 0.13$. However, only about 45\% of stars in this age range are expected to host a disc \citep{Rigliaco2025} and could therefore be represented in our sample. If the observed shortage of large grains is indeed related to the ongoing formation of giant planets, the product of these two fractions would imply an overall fraction of solar-type stars hosting a giant planet of $0.27\pm0.06$ in this age range. This value is broadly consistent with estimates of the fraction of giant planets around solar-type stars in nearby young associations \citep{Gratton2024, Gratton2025}, which may represent the most likely progeny of the low-density star-forming regions considered in this study. In these papers it was found that the fraction of  giant planets around solar-type stars in nearby young associations depends on the mass and/or age of the association, with values in the range from 0.2 to 1. However, this agreement cannot be taken as evidence for a direct connection. Filtration may be caused by other effects and possibly also by cores that never become giant planets. Our result only shows that the frequency of disc exhibiting filtration is consistent with the hypothesis that the observed depletion of large grains could be associated with ongoing giant-planet formation.

\begin{figure*}[htb!]
\sidecaption
    \includegraphics[width=12cm]{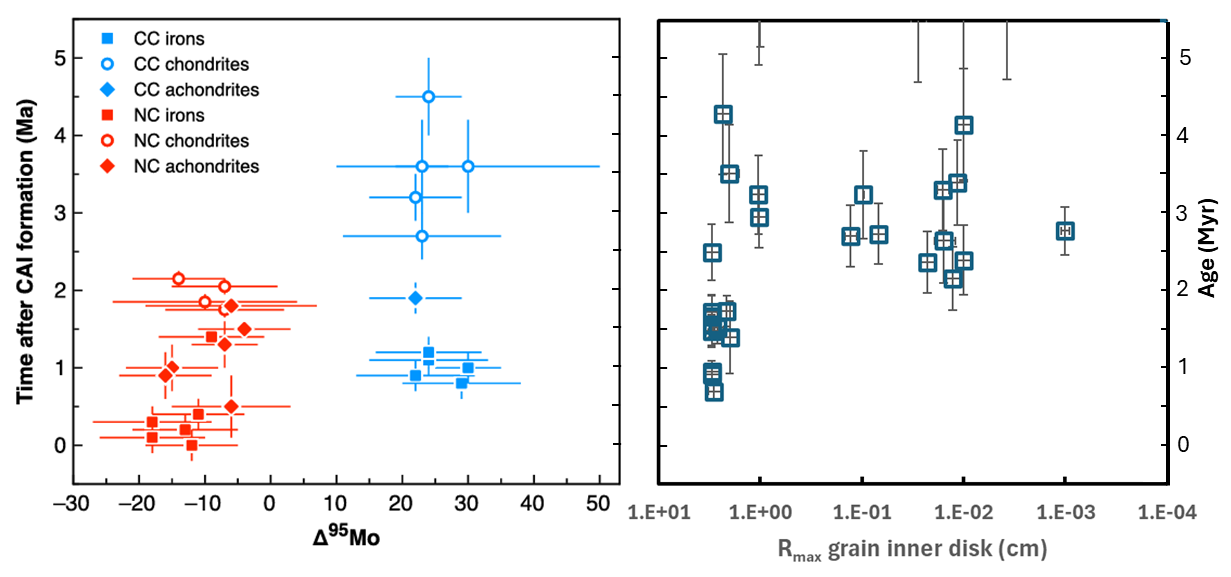}
    \caption{Left: Plot of meteorite accretion ages versus the $^{95}$Mo abundance anomaly $\Delta ^{95}$Mo. The NC–CC dichotomy combined with the accretion ages demonstrates that NC and CC meteorites derive from two spatially separated and contemporaneous reservoirs. The two reservoirs were most plausibly separated by Jupiter. From \citet{Kleine2020}. Right: Trends with age for the maximum radius of grains in the inner disc of stars with masses similar to the Sun from this study.  }
    \label{fig:solar_system_comp}
\end{figure*}

\subsection{Comparison with the Solar System}

As discussed above, our sample is biassed towards bright and long-lived discs \citep{vanderMarel2023}. Should future studies confirm that the results obtained for this sample are representative of the wider population of young solar-type stars — for instance, because giant planets form predominantly in long-lived discs — a minimum formation timescale of $\sim 2$\, Myr for giant-planet cores with masses greater than $>30$\, \MEarth would be in good agreement with the core-accretion paradigm (e.g., \citealt{Pollack1996, Mordasini2012, Armitage2020}). \citet{Kruijer2017} and \citet{Kleine2020} discussed evidence for the Solar System and about the formation epoch of Jupiter, mainly based on the isotopic composition of meteorites. An interesting datum is the ratio between the abundances of the isotopes of molibdenum, in particular the offset (in parts-per-ten-thousand) of the abundance of $^{95}$Mo normalized to that of $^{98}$Mo with respect to the Earth value ($\Delta ^{95}$Mo), since these two isotopes have different nucleosynthesis. In their scenario, the separation between NC (in the inner part of the disc, characterised by negative values of $\Delta ^{95}$Mo) and CC in the outer part, characterised by positive values of $\Delta ^{95}$Mo) occurred about 1\, Myr after the origin of the Solar System (as described by the formation of calcium-aluminium inclusion, CAI), possibly due to the formation of the early core of Jupiter with a mass of about 20\, M$_{\rm Earth}$ by pebble accretion \citep{Alibert2018}. There is a clear analogy with the evolution of the maximum size of the grains in the inner disc as derived in this paper (see Fig.\, \ref{fig:solar_system_comp}) suggesting a similar cause. 
Interestingly, three of the stars in our sample with suspected, but not yet confirmed, planetary companions -- LkCa 15, WRAY 15--1880, and CQ Tau -- are among those showing a depletion of large grains in their inner discs. A similar comparison was discussed in the previous section in the context of the study by \citealt{Ruzza2026}. The same behaviour is not observed for the two youngest stars with proposed planetary companions, IM Lup and GK Tau, nor for PDS 66, which, with an age of $4.3\pm0.8$,Myr, is the oldest star in our sample within the 2--5,Myr age range.
The separation between the inner and outer reservoirs of grains appears to have occurred earlier in the Solar System than in the discs analysed in our study. The discrepancy may result from the uncertainties associated with the different methods used to constrain these timescales, or alternatively may reflect a genuine difference between the early evolution of the Solar System and that of other stars of similar mass considered here. The latter possibility could be related to the fact that the stars in our sample typically retain their protoplanetary discs for relatively long timescales.
It is also possible that separation between the two reservoirs, as inferred from their distinct chemical compositions, occurs earlier than the emergence of a measurable difference in the grain-size distribution. The filtering effect of the planetary core may initially be masked by efficient dust coagulation during the early stages following its onset, when the inner disc is still sufficiently dense to promote rapid grain growth.

However, very recently \citet{Bryson2026} proposed a more complex scenario for the Solar System based on a refined thermal evolution model to calculate the formation ages of meteorite parent planetesimals. In this scenario, at an early time ($\sim 0.9$\, Myr after CAI formation), the disc is divided by a barrier feature that creates an inner reservoir and an outer reservoir composed of CC material. CAIs are localised at the position of this barrier. At $\sim 0.95$\, Myr after CAI formation, the nature of the barrier changes, becoming semi-permeable. After this time, this feature has three effects: it transports NC material outward and CC material inward; it causes planetesimals to form on both sides; and it moves CAIs overwhelmingly outwards. By $\sim 2$\, Myr after CAI formation, more NC and CC material has crossed the feature and more planetesimals have formed. Between $\sim 2-4$\, Myr after CAI formation, the NC material no longer mixes into the CC reservoir and the distal CC material starts mixing into the innermost CC reservoir. More planetesimals form during this period in the CC reservoir, and the NC reservoir may have dissipated by this time. The proposed scenario more closely reproduces our timing for the different phases of disc evolution.

\citet{Kruijer2017} proposed that the next growth of the Jupiter core up to a mass of about 50\, M$_{\rm Earth}$ was slow, possibly because only planetesimal accretion occurs in this phase \citep{Alibert2018}, and the gas accretion phase only occurred after more than 4\, Myr from CAI formation. In addition, even after this phase, very young giant planets may still be embedded in thick circum-planetary discs \citep{Lubow1999} that prevent their detection for some time \citep{Maio2025}, as predicted by models for the formation of giant planets \citep{DAngelo2018}, and agreeing with the observation that known accreting planets have ages of about 6\, Myr, and several of the programme stars younger than this have been searched for planets emitting in H$\alpha$ without any detection. They include MWC 758 (2.6\, Myr), LkCa 15 (2.8\, Myr, but see \citealt{Close2025} for a possible detection), RX\, J1615.3--3255 (2.7\, Myr), WRAY 15--1880 (3.4\, Myr), and PDS 66 (4.3\, Myr). If this reasoning is correct, the observation of accretion is possible only in the late stages of the planet growth (and possibly in systems seen from a favourable angle, see \citealt{Close2025}).

\section{Conclusions}
\label{sect:conclusions}

In this study we exploited the dynamical masses obtained from the kinematics of gas in discs around 33 young solar type stars (mass in the range 0.7--1.4\, \Msun) to derive the evolution of basic properties of the discs and the accretion rate onto the star. We did this by fitting the SED and the ALMA dust continuum images with a simple parametric model of the emission by the star and the disc. The main quantities we derived are the radius and effective temperature of the stars, the accretion rate, the dust mass and extension of the inner and outer regions of the disc, and the maximum radius of the grains in the inner disc. We combined stellar radii and masses with predictions of the evolutionary models of \citet{Baraffe2015} to derive the age of the individual stars. The sample is obviously biassed with respect to the presence of a relevant disc that is required to extract the stellar masses \citep{vanderMarel2023}. 

We highlight that the individual parameters derived from our analysis are closely related to observables. The accretion rate is correlated to the UV flux, the stellar effective temperature is related to the colour and the stellar radius to the absolute visual magnitude in the optical (this last, modulated by the effective temperature), the inner radius of the inner disc to thermal IR colours (such as $W3-W4$), the dust mass of the inner and outer parts of the disc to the central and total emission in the ALMA images, respectively. When emission at the centre of the ALMA image is very low, the dust mass of the inner disc is actually constrained by the fit of the SED.

Our analysis allowed us to find trends of the quantities mentioned above with the age of the star. As expected, we found that the accretion rate and dust mass of the inner disc evolve with time, with a typical reduction of three orders of magnitude in the range of ages 1--8\, Myr. The evolution of the dust mass in the outer portion of the disc is less obvious; however, it becomes more clear when combined with the reduction in the fraction of stars hosting discs with age.

Assuming a gas-to-dust ratio of 100, the mass of the inner disc explains quite well the observed accretion rates for stars younger than 2\, Myr for a viscous disc with $\alpha\sim 0.001$. If we still assume a gas-to-dust ratio of 100, we find much lower dust mass in the inner disc (60\%) of discs with ages in the range 2--5\, Myr than that required to sustain the observed accretion rates (unless we assume an unrealistic value of $\alpha\sim 1$). The lower dust content is characterised by a depletion of large dust grains, which is a direct consequence of the low central fluxes in the ALMA dust continuum images for these discs. Interestingly, the three discs with strong asymmetries in the ALMA images in our sample (see Sect.~\ref{sect:strong_asymmetry}) have ages of about 2.4\, Myr. Such asymmetries might be related to the presence of planets inducing dust traps in the disc. 

The relative depletion of large grains in 9 out of 15 of the inner discs around stars with ages in the range 2--5\, Myr suggests the presence of some mechanism stopping the inward drift of large grains. One possibility for the depletion of large grains is the progressive growth of the cores of giant planets in agreement with the current scenarios for the formation of Jupiter \citep{Pollack1996, Zhang2018, Alibert2018, Kleine2020}. Indeed, the presence of giant planets has been proposed for a few of these cases (CQ Tau: \citealt{Marsh1993}; LkCa 15: \citealt{Sallum2015, Close2025}; WRAY 15--1880: \citealt{Rigliaco2026}). The observed depletion in more than half of the discs older than 2 Myr may therefore point to the presence of giant planets or planetary cores more massive than 30 \MEarth, although the exact mass threshold remains uncertain. On a speculative ground, if we might assume that only long-living discs (those considered in the present study) may form giant planets, and we take into account that only about 45\% of the stars in this age range still host a disc, the overall fraction of solar type stars hosting a giant planet in this age range is $0.27\pm 0.06$. The derived fraction of solar type stars hosting a giant planet is not far from the estimates of the fraction of giant planets around solar-type stars in nearby young associations \citep{Gratton2024, Gratton2025} that are the most likely progeny of the low density star forming regions considered in this study. 

Of course, we might expect that the evolution of individual objects might differ from this broad scenario. For instance, we expect that (later) formation of giant planets will be difficult around stars younger than 2\, Myr lacking prominent discs (some 40\% of total; see e.g. \citealt{Rigliaco2025}). Also, all the discs considered in this study are in low density regions. Discs in higher density regions, such as the Orion Nebula cluster, are known to have different properties, being, for example, more compact \citep{Eisner2018}. In general, the formation of giant planets is likely sensitive to the environment \citep{Gratton2024, Gratton2025} and investigating these aspects requires further studies. However, a consistent scenario seems to emerge where the formation of giant planets around solar type stars occurs between 1 and 6\, Myr after the star formation.\\

\noindent
{\bf{Data availability}}\\ 
\noindent
The catalog table associated with this article is available at the CDS via https://cdsarc.cds.unistra.fr/viz-bin/cat/J/A+A/707/A61.
The Appendices C, D and E are available via \href{https://doi.org/10.5281/zenodo.22792373}{this link}

\begin{acknowledgements}
We thank the anonymous referee for the detailed report that has improved the quality of the paper. 
This paper makes use of the following ALMA data: ADS/JAO.ALMA\#2015.1.01083.S. ALMA is a partnership of ESO (representing its member states), NSF (USA) and NINS (Japan), together with NRC (Canada), NSTC and ASIAA (Taiwan), and KASI (Republic of Korea), in cooperation with the Republic of Chile. The Joint ALMA Observatory is operated by ESO, AUI/NRAO and NAOJ.
E.R. acknowledges support from PRIN-MUR 2022 20228JPA3A “The path to star and planet formation in the JWST era (PATH).” The work was also partially funded with an INAF "Mini-Grant" RF 2022. 
E.R. gratefully acknowledges support from the “Programma di Ricerca Fondamentale INAF 2023” of the Italian National Institute of Astrophysics (INAF Large Grant 2023 “NextSTEPS”). 
\end{acknowledgements}

\bibliographystyle{aa} 
\bibliography{biblio} 

\begin{appendix}
\relax 

\section{Input data }
\label{app:input_data}

The input parameters for the stars are provided in Tables\, \ref{tab:uv_data}, \ref{tab:input_data}, and  \ref{tab:alma_data}. The first table gives the coordinates of the stars and information about space UV data available. The second table includes the parallax, the spectral type (from SIMBAD data base), the $B_P-R_P$ colour from {\it Gaia}\, DR3, the interstellar absorption ($A_V$) that is obtained by considering the relation between spectral type and intrinsic colour of \citet{Pecaut2013}, the dynamic mass of the star with its reference, and the inclination $i$ and position angle $PA$ of the disc with their references. The third table gives the id of the ALMA high resolution observations used in the analysis, the band, the version of the {\tt CASA} pipeline used in the data reduction, the continuum sensitivity, the space resolution, the total fluxes in the two bands, and the intensity at the centre of the image.

In the following, we give some additional notes for some individual objects. Some of the stars considered in the final sample are likely unresolved close binaries (separation $<0.1^{\prime\prime}$). In these cases, the mass given by gas kinematics is the sum of the mass of the two components. In our analysis, we used the mass value that is appropriate for the primary and did not consider the contribution of the secondary to the SED because this is negligible in all cases.\\
\paragraph{IQ Tau:} \citet{Semenov2024} obtained 0.648 \Msun.
\paragraph{RU Lup:}  the high value of the Renormalised Unit Weight Error (RUWE) parameter from {\it Gaia} indicates that this star is a close binary. Taking into account the available data (photometry, RUWE, radial velocities from {\it Gaia}, and lack of detection of a visual companion), the procedure described in \citet{Gratton2023} yields the best astrometric solution with $m_A$ = 1.14 \, \Msun, $m_B$ = 0.44 \, \Msun, $a=2.4$\, au.
\paragraph{GM Aur:} the high value of RUWE indicates that this star is a close binary. Using the same method described for RU Lup, we found that this indicates the presence of a small mass companion of $0.018\pm 0.006$\, \Msun at 0.24 au. This separation is compatible with the value of $a_{\rm min}$(inner) we derived for this star, suggesting that the dust disc is circumbinary.
\paragraph{CI Tau:} \citet{Semenov2024} obtained 1.004\, \Msun.
\paragraph{RX\, J1615.3--3255:} \citet{Longarini2025} obtained $1.105\pm 0.012$ \Msun. The high value of RUWE indicates that this star is a close binary but the secondary is a low mass star. Our astrometric solution gives a mass of 0.13\, \Msun. The difference with respect to the value we adopted is due to the companion.
\paragraph{GK Tau:} \citet{Long2019} obtain $i=40.2$\ and $PA=119.9$\, deg based on their analysis of ALMA data.
\paragraph{CQ Tau:} \citet{Izquierdo2025} obtained 1.4\, \Msun. The high value of RUWE indicates that this star is a close binary. The astrometric solution using the same method described for RU Lup yields $m_A=1.07$\, \Msun, $m_B=0.33$\, \Msun, $a$=1.3 au.
\paragraph{RX\, J1852.3-3700:} \citet{Longarini2025} obtained a mass of 1.022\, \Msun. The high value of RUWE indicates that this star is a close binary but the secondary is a low mass star. Our astrometric solution gives a mass of 0.09\, \Msun for the secondary.
\paragraph{2MASS J16120668--3010270:} \citet{Zhang2025} gives 0.51\, \Msun.
\paragraph{V1094 Sco:} \citet{Braun2021} obtained 1.06\, \Msun. For comparison, \citet{Zhang2025} proposed a mass of 0.82\, \Msun and \citet{Trapman2025} of 1.10\, \Msun. The high value of RUWE suggests the presence of a companion. Using the same method described for RU Lup, we found that this companion should have a mass of 0.24\, \Msun and is at 1.7\, au.

\begin{table*}[!h]
\caption{Space UV data.}
\centering
\small
\begin{tabular}{lccccccccc}
\hline
\hline
Star & RA	&	Dec	& \multicolumn{2}{c}{HST}  &\multicolumn{2}{c}{Galex} & \multicolumn{3}{c}{XMM} \\	
&(deg)	&	(deg)	&		COS	&	STIS 	&	FUV	&	NUV	 &	UVW2	&	UVM &	UVW1 \\
\hline
BP Tau	&64.81602	&	29.10736	&	Y	&	Y	&	Y	&	Y	&	Y	&		&	Y	\\
IP Tau	&66.23788	&	27.19892 	&	Y	&		&		&		&		&		&		\\
IQ Tau	&67.46485	&	26.11237	&		&		&		&		&	Y	&		&		\\
GK Tau	&68.39405	&	24.35154	&		&		&	Y	&	Y	&	Y	&	Y	&	Y	\\
CI Tau	&68.46677	&	22.84162	&		&		&	Y	&	Y	&	Y	&	Y	&		\\
LkCa 15	&69.82418	&	22.35086	&	Y	&	Y	&		&		&		&	Y	&		\\
DR Tau	&71.77592	&	16.97850	&	Y	&	Y	&	Y	&	Y	&		&	Y	&	Y	\\
DS Tau	&71.95251	&	29.41967	&		&	Y	&		&		&		&		&		\\
GM Aur	&73.79578	&	30.36638	&	Y	&	Y	&	Y	&	Y	&	Y	&	Y	&	Y	\\
V836 Tau&75.77751	&	25.38870	&		&		&	Y	&	Y	&	Y	&		&		\\
MWC 758	&82.61472	&	25.33240	&	Y	&		&		&		&	Y	&		&		\\
CQ Tau	&83.99363	&	24.74824	&		&		&	Y	&	Y	&	Y	&		&		\\
SY Cha	&164.12614	&	-77.19427	&	Y	&	Y	&	Y	&	Y	&	Y	&		&		\\
PDS 66	&200.53094	&	-69.63682	&		&	Y	&	Y	&		&	Y	&	Y	&	Y	\\
PDS 70	&212.04213	&	-41.39804	&		&	Y	&		&		&		&	Y	&	Y	\\
Sz 65	&234.86564	&	-34.77155	&		&	Y	&		&	Y	&		&		&		\\
GW Lup	&236.68630	&	-34.51001	&		&	Y	&		&		&		&		&		\\
GQ Lup	&237.30036	&	-35.65151	&	Y	&		&		&		&		&	Y	&		\\
Sz 77	&237.94559	&	-35.94569	&		&	Y	&		&		&		&		&		\\
RX\, J1556.1--3655	&239.00868	&	-36.92462	&	Y	&	Y	&		&		&		&		&		\\
IM Lup	&239.03829	&	-37.93514	&	Y	&	Y	&		&		&		&		&		\\
RU Lup	&239.17623	&	-37.82107	&		&	Y	&		&		&		&		&		\\
Sz 129	&239.81857	&	-41.95296	&	Y	&	Y	&		&		&		&		&		\\
RX\, J1604.3--2130A	&241.09017	&	-21.50804	&		&		&		&		&		&		&		\\
V1094 Sco	&242.15068	&	-39.38412	&
		&		&		&		&		&		&		\\
2MASS J16090075--1908526	&242.25312	&	-19.14808	&		&		&		&		&		&		&		\\
2MASS J16120668--3010270	&243.02777	&	-30.17431	&		&		&		&		&		&		&		\\
2MASS J16124373--3815031	&243.18224	&	-38.25095	&		&	Y	&		&		&		&		&		\\
RX\, J1615.3--3255	&243.83427	&	-32.91819	&		&		&		&		&		&		&		\\
2MASS J16202863--2442087	&245.11928	&	-24.70251	&		&		&		&		&		&		&		\\
AS 209	&252.31373	&	-14.36917	&		&		&		&	Y	&	Y	&		&	Y	\\
WRAY15--1880	&280.74160	&	-35.54535	&	Y	&	Y	&		&		&	Y	&		&		\\
RX\, J1852.3-3700	&283.07211	&	-37.00345	&	Y	&	Y	&	Y	&	Y	&	Y	&		&		\\
\hline
\end{tabular}
\label{tab:uv_data}
\end{table*}

\begin{table*}[!htb]
\caption{Input parameters.}
\centering
\small
\begin{tabular}{lccccccccc}
\hline
\hline
Star& Parallax &Sp.T.&$B_P-R_P$&$A_V$&$M_\star$&Ref.&$i$&$PA$&Ref.\\
&(mas)&&(mag)&(mag)&(\Msun)&&(deg)&(deg)&\\
\hline
BP Tau	&7.849	&K5/7Ve	&0.106	&0.222	&1.100	&5	&37.81	&151.10	&10\\
IP Tau	&7.729	&M0	&0.183	&0.386	&0.800	&1	&47.00	&173.50	&2\\
IQ Tau	&7.604	&M0.5	&0.504	&1.061	&0.740	&5 	&62.12	&42.38	&2\\
GK Tau	&7.743	&K7:Ve	&0.434	&0.915	&	0.730	&5	&68.50	&134.00	&12\\
CI Tau	&6.238	&K4IVe	&0.789	&1.661	&0.900	&5	&49.99	&11.22	&2\\
LkCa 15	&6.427	&K5	&0.132	&0.278	&1.118	&8	&50.59	&61.57	&12\\
DR Tau	&5.182	&K5	&0.191	&0.403	&1.180	&1	&5.40	&3.40	&2\\
DS Tau	&6.315	&K4	&0.363	&0.765	&1.080	&1	&64.00	&160.00	&2\\
GM Aur	&6.325	&K3	&0.411	&0.865	&1.100	&3  &53.65	&56.95	&12\\
V836 Tau&5.988	&K7	&0.410	&0.864	&0.920	&1	&64.00	&307.00	&9\\
MWC 758	&6.416	&A8	&0.105	&0.220	&1.400	&7	&7.27	&76.17	&12\\
CQ Tau	&6.695	&F5	&0.571	&1.203	&1.069&	12	&33.65	&54.50 &12\\	
SY Cha	&5.533	&K5	&0.529	&1.114	&0.812	&8	&51.65	&165.77	&12\\
PDS 66	&10.215	&K1	&--0.031	&0.000	&1.299	&8	&32.02	&8.91 &12\\
PDS 70	&8.898	&K7	&--0.027	&0.000	&0.760	&11	&51.00	&158.20	&11\\
Sz 65	&6.516	&K7	&0.267	&0.562	&0.770	&4	&63.00	&110.00	&4\\
GW Lup	&6.443	&M1.5e	&0.245	&0.517	&0.770	&6	&48.00	&36.00	&6\\
GQ Lup	&6.489	&K7	&0.043	&0.090	&0.960	&1	&60.20	&167.90	&1\\
Sz 77	&6.441	&K5.5	&0.316	&0.665	&0.790	&6 	&33.00	&108.00	&6\\
RX\, J1556.1--3655	&6.331	&M1	&--0.209	&0.000	&0.750	&1	&53.50	&55.60	&1\\
IM Lup	&6.417	&M0	&--0.065	&0.000	&1.100	&3	&53.30	&315.60	&12\\
RU Lup	&6.349	&K7/M0	&--0.381	&0.000	&1.000	&1 	&3.31	&163.76	&1\\
Sz 129	&6.245	&K7	&0.187	&0.394	&0.850	&1	&31.74	&154.94	&1\\
RX\, J1604.3--2130A	&6.882	&K2	&0.525	&1.106	&1.290	&7	&8.72	&123.24	&12\\
V1094 Sco	&6.462	&K6	&0.480	&1.010	&0.869 &12	&49.20	&107.40	&1\\
2MASS J16090075--1908526	&7.278	&M1.0V	&0.099	&0.208	&0.900	&6	&49.00&325.00	&6\\
2MASS J16120668--3010270	&7.568	&M0.5	&--0.020	&0.000	&0.730	&6 	&36.00	&45.00	&6\\
2MASS J16124373--3815031	&6.256	&M1	&0.133	&0.280	&0.770	&1	&54.00	&22.99	&1\\
RX\, J1615.3--3255	&6.362	&K5	&0.209	&0.440	&0.972	&12	&47.10	&146.14	&12\\
2MASS J16202863--2442087	&6.490	&M2	&0.537	&1.131	&0.700	&6	&40.00	&188.00	&6\\
AS 209	&8.248	&K4	&0.643	&1.354	&1.200	&3	&34.88	&85.76	&12\\
WRAY15--1880 &6.621	&K2	&0.429	&0.903	&1.042	&8	&39.22	&26.35	&8\\
WRAY15--1880 &6.621	&K2	&0.429	&0.903	&1.042	&8	&39.22	&26.35	&8\\
RX\, J1852.3-3700	&6.799	&K5	&0.064	&0.135	&0.930		&3&2.50	&117.61	&12\\
\hline
\end{tabular}
\tablebib{
(1) \citet{Braun2021};
(2) \citet{Long2018};
(3) \citet{Bosman2021};
(4) \citet{Miley2024};
(5) \citet{Simon2019};
(6) \citet{Trapman2025};
(7) \citet{Izquierdo2025};
(8) \citet{Longarini2025};
(9) \citet{Najita2008};
(10) \citet{Aizawa2020};
(11) \citet{Keppler2019};
(12) This paper.
}
\label{tab:input_data}
\end{table*}

\begin{table*}[!htb]
\caption{ALMA data taken from the ALMA archive. }
\centering
\small
\begin{tabular}{lcccccccc}
\hline
\hline
Star&Program&Band&{\tt CASA}&Cont. Sens.&Res.& F$_{870}$ & F$_{1330}$&Center\\
&&&Version&(mJy\, beam$^{-1}$)&(arcsec)&(mJy)&(mJy)&(mJy\, beam$^{-1}$)\\
\hline
BP Tau	&	2019.1.00607.S	&	6	& 5.1.1-5&	0.0123	&	0.024	&	130.0	&	58.6	&	51.7	\\
IP Tau	&	2016.1.01164.S	&	6	& 5.1.1-5&	0.0750	&	0.112	&	32.0	&	8.8	&	6.0	\\
IQ Tau	&	2016.1.01164.S	&	6	& 5.1.1-5&	0.0765	&	0.117	&	87.0	&	61.9	&	8.0	\\
GK Tau	&	2018.1.00771.S	&	6	& 5.6.1-8&	0.0686	&	0.144	&	...	&	21.0	&	2.6	\\
CI Tau	&	2016.1.01370.S	&	6	& 5.1.1-5&	0.0159	&	0.035	&	324.0	&	190.0	&	23.8	\\
LkCa 15	&	2018.1.00945.S	&	6	& 5.4.0-70&	0.0196	&	0.028	&	407.1	&	147.0	&	--2.8	\\
DR Tau	&	2022.1.01365.S	&	6	& 6.4.1.12&	0.0157	&	0.022	&	424.0	&	130.0	&	29.0	\\
DS Tau	&	2016.1.01164.S	&	6	& 5.1.1-5&	0.0440	&	0.115	&	40.5	&	16.5	&	7.2	\\
GM Aur	&	2017.1.01151.S	&	6	& 5.1.1-5&	0.0147	&	0.024	&	...	&	213.0	&	0.2	\\
V836 Tau	&	2016.1.01164.S	&	6	& 5.1.1-5&	0.0449	&	0.111	&	70.6	&	31.0	&	22.4	\\
MWC 758	&	2017.1.00940.S	&	6	& 5.1.1-5&	0.0166	&	0.022	&	700.0	&	226.3	&	1.3	\\
CQ Tau	&	2017.1.01404.S	&	6	& 6.2.1-7&	0.0149	&	0.056	&	581.7	&	...	&	1.5	\\
SY Cha	&	2018.1.00689.S	&	6	& 6.2.1-7&	0.0168	&	0.024	&	158.4	&	...	&	1.59	\\
PDS 66	&	2017.1.01167.S	&	6	& 6.2.1-7&	0.0291	&	0.056	&	336.1	&	...	&	106.0	\\
PDS 70	&	2018.A.00030.S	&	7	& 5.6.1-8&	0.0287	&	0.027	&	176.0	&	...	&	0.2	\\
Sz 65	&	2018.1.00271.S	&	6	& 5.1.1-5&	0.0168	&	0.016	&	64.5	&	29.9	&	31.5	\\
GW Lup	&	2016.1.00484.L	&	6	& 5.1.1-5&	0.0184	&	0.024	&	166.0	&	69.2	&	14.1	\\
GQ Lup	&	2022.1.01302.S	&	7	& 6.4.1.12&	0.0569	&	0.098	&	...	&	42.0	&	13.0	\\
Sz 77	&	2022.1.00154.S	&	6	& 6.5.4-9&	0.0325	&	0.032	&	...	&	1.9	&	22.8	\\
RX\, J1556.1--3655	&	2022.1.01302.S	& 	6	& 6.4.1-12&	0.0369	&	0.099	&	...	&	23.5	&	6.1	\\
IM Lup	&	2016.1.00484.L	&	6	& 6.5.4.9&	0.0148	&	0.023	&	506.0	&	...	&	39.2	\\
RU Lup	&	2016.1.00484.L	&	6	& 5.1.1-5&	0.0161	&	0.022	&	...	&	28.2	&	10.9	\\
Sz 129	&	2018.1.01054.S	&	6	& 5.4.0-70&	0.0330	&	0.021	&	181.0	&	75.9	&	0.2	\\
RX\, J1604.3--2130A	&	2018.1.01255.S	&	6	& 6.2.1-7&	0.0167	&	0.029	&	274.0	&	...	&	-0.3	\\
V1094 Sco	&	2017.1.01167.S	&	6	& 6.2.1-7&	0.0252	&	0.048	&	...	&	263.4	&	5.4	\\
2MASS J16090075--1908526	&	2017.1.01167.S	&	6	& 5.6.1-8&	0.0298	&	0.041	&	47.3	&	12.3	&	23.2	\\
2MASS J16120668--3010270	&	2022.1.00646.S	&	6	& 6.4.1-12&	0.0159	&	0.034	&	...	&	10.4	&	2.9	\\
2MASS J16124373-3815031	&	2022.1.00154.S	&	6	& 6.4.1-12&	0.0311	&	0.030	&	29.9	&	11.8	&	25.5	\\
RX\, J1615.3--3255	&	2016.1.01286.S	&	7	& 6.5.4-9&	0.0341	&	0.063	&	386.0	&	...	&	11.1	\\
2MASS J16202863--2442087	&	2021.1.00128.L	&	6& 	6.2.1-7&	0.0459	&	0.138	&	...	&	1.5	&	0.1	\\
AS 209	&	2016.1.00484.L	&	6	& 6.5.4-9&	0.0160	&	0.033	&	598.0	&	300.0	&	29.6	\\
WRAY15--1880	&	2021.1.01123.L	&	7& 	6.5.4-9&	0.0564	&	0.094	&	135.7	&	...	&	1.5	\\
WRAY15--1880	&	2015.1.01083.S	&	7& 	4.7.38335&	0.0601	&	0.166	&	135.7	&	...	&	1.2	\\
RX\, J1852.3-3700	&	2018.1.00689.S	&	6& 	6.2.1-7&	0.0116	&	0.021	&	150.9	&	...	&	0.4	\\
\hline
\end{tabular}
\label{tab:alma_data}
\end{table*}

\begin{table*}[htb]
\caption{Output parameters. }
\tiny
\centering
\begin{tabular}{lccccccccc}
\hline
\hline
Star	&	$\chi^2$(SED)	&	$\chi^2$(ALMA)	&	\teff	&	$R_*$	& $\log{\dot{M}}$	& $a_{\rm min,inner}$	&	$a_{\rm max,inner}$	&	$M_{\rm inner}$	&	$M_{\rm outer}$		\\
&&& (K) & (\RSun) &(M$_\odot\,yr^{-1}$) & (au) & (au) & ($10^{-5}$\, \Msun) & ($10^{-3}$\, \Msun)\\
\hline
BP Tau	&	41.1	&	15.9	&	4091	$\pm$	30	&	1.69	$\pm$	0.10	&	--8.84	$\pm$	0.10	&	0.068		$\pm$	0.003	&	9.0	&	9.1$\pm$0.4	&	8.6$\pm$0.5		\\
IP Tau	&	14.7	&	7.4	&	3850	$\pm$	27	&	1.46	$\pm$	0.05	&	--9.60	$\pm$	0.10	&	0.053		$\pm$	0.003	&	2.8	&	3.5$\pm$0.2	&	1.3$\pm$0.1		\\
IQ Tau	&	19.4	&	21.6	&	3417	$\pm$	25	&	1.91	$\pm$	0.09	&	--8.76	$\pm$	0.10	&		0.022	$^a$	&	1.7	&	1.8$\pm$0.1	&	12$\pm$1		\\
GK Tau	&	33.6	&	24.3	&	4009	$\pm$	30	&	1.84	$\pm$	0.10	&	--8.71	$\pm$	0.11	&	0.049		$\pm$	0.003	&	2.3	&	6.0$\pm$0.4	&	4.0$\pm$0.2		\\
CI Tau	&	16.7	&	91.7	&	4487	$\pm$	28	&	1.72	$\pm$	0.09	&	--8.12	$\pm$	0.11	&	0.047		$\pm$	0.003	&	2.7	&	3.6$\pm$0.2	&	12$\pm$1		\\
LkCa 15	&	17.6	&	14.0	&	4167	$\pm$	28	&	1.82	$\pm$	0.07	&	--9.40	$\pm$	0.11	&	0.122		$\pm$	0.005	&	2.4	&	26$\pm$2	&	28$\pm$2		\\
DR Tau	&	25.5	&	36.5	&	3901	$\pm$	30	&	3.04	$\pm$	0.14	&	--7.67	$\pm$	0.10	&	0.047		$^a$	&	3.67	&	37$\pm$7	&	25$\pm$1		\\
DS Tau	&	6.5	&	9.18	&	4070	$\pm$	27	&	1.75	$\pm$	0.08	&	--8.15	$\pm$	0.08	&	0.037		$\pm$	0.001	&	2.5	&	1.6$\pm$0.1	&	4.9$\pm$0.3		\\
GM Aur	&	100.1	&	134.1	&	4196	$\pm$	28	&	1.96	$\pm$	0.11	&	--8.27	$\pm$	0.11	&		1.431	$\pm$	0.088	&	2.7	&	74$\pm$4	&		16$\pm$1	\\
V836 Tau	&	14.2	&	16.3	&	3869	$\pm$	29	&	1.70	$\pm$	0.08	&	--9.74	$\pm$	0.09	&	0.055		$\pm$	0.004	&	16	&	1.6$\pm$1	&	12$\pm$1		\\
MWC 758	&	116.1	&	612.8	&	7686	$\pm$	28	&	2.06	$\pm$	0.11	&	--7.86	$\pm$	0.11	&	0.130		$^a$	&	10.2	&	130$\pm$10	&		7.3$\pm$0.4	\\
CQ Tau	&	37.4	&	191.4	&	5583	$\pm$	30	&	1.90	$\pm$	0.10	&	--7.36	$\pm$	0.12	&		0.062	$^a$	&	10.2	&	390$\pm$20	&	22$\pm$1		\\
SY Cha	&	26.5	&	20.9	&	4046	$\pm$	26	&	1.68	$\pm$	0.09	&	--8.20	$\pm$	0.08	&	0.123		$\pm$	0.007	&	13.2	&	63$\pm$3	&	35$\pm$2		\\
PDS 66	&	29.2	&	31.5	&	4286	$\pm$	30	&	1.73	$\pm$	0.08	&	--9.04	$\pm$	0.11	&	0.071		$\pm$	0.004	&	2.5	&	2.1$\pm$0.1	&	7.7$\pm$0.4		\\
PDS 70	&	22.1	&	12.7	&	4106	$\pm$	29	&	1.25	$\pm$	0.05	&	--10.66	$\pm$	0.12	&	0.122		$\pm$	0.007	&	27	&	59$\pm$3	&	9.2$\pm$0.5		\\
Sz 65	&	34.4	&	16.8	&	4025	$\pm$	29	&	1.95	$\pm$	0.10	&	--9.58	$\pm$	0.12	&	0.031		$^a$	&	2.4	&	1.2$\pm$0.1	&	5.3$\pm$0.3		\\
GW Lup	&	67.8	&	111.2	&	3561	$\pm$	28	&	1.65	$\pm$	0.10	&	--9.67	$\pm$	0.12	&	0.106		$\pm$	0.006	&	2.3	&	3.0$\pm$0.2	&	10$\pm$1		\\
GQ Lup	&	28.9	&	31.8	&	3942	$\pm$	29	&	2.44	$\pm$	0.11	&	--8.40	$\pm$	0.10	&	0.038		$^a$	&	3.0	&	4.9$\pm$0.3	&	17$\pm$1		\\
Sz 77	&	57.4	&	6.1	&	4019	$\pm$	28	&	1.87	$\pm$	0.09	&	--8.77	$\pm$	0.11	&	0.052		$\pm$	0.003	&	2.3	&	1.4$\pm$0.1	&	5.7$\pm$0.3		\\
RX\, J1556.1--3655	&	30.5	&	20.55	&	3627	$\pm$	30	&	1.48	$\pm$	0.07	&	--9.17	$\pm$	0.09	&	0.130		$\pm$	0.008	&	1.5	& 1.5$\pm$0.1	&	7.7$\pm$0.4		\\
IM Lup	&	36.5	&	97.6	&	3814	$\pm$	26	&	2.65	$\pm$	0.13	&	--8.81	$\pm$	0.12	&	0.116		$\pm$	0.006	&	2.9	&	3.2$\pm$0.2	&	5.6$\pm$0.3		\\
RU Lup	&	25.1	&	166.1	&	3902	$\pm$	26	&	2.16	$\pm$	0.13	&	--7.62	$\pm$	0.10	&		0.035	$^a$	&	9.1	&	190$\pm$10	&	15$\pm$1		\\
Sz 129	&	33.8	&	9.0	&	4097	$\pm$	28	&	1.42	$\pm$	0.07	&	--9.07	$\pm$	0.12	&	0.034		$\pm$	0.002	&	7.2	&	11$\pm$1	&	6.8$\pm$0.4		\\
RX\, J1604.3--2130A	&	25.5	&	59.0	&	4551	$\pm$	29	&	1.55	$\pm$	0.08	&	--10.15	$\pm$	0.12	&	0.032		$^a$	&	5.1	&	5.1$\pm$0.3	&	26$\pm$1		\\
V1094 Sco	&	33.1	&	30.6	&	4385	$\pm$	28	&	1.28	$\pm$	0.06	&	--11.31	$\pm$	0.11	&		0.025	$^a$	&	13	&	5.1$\pm$0.3	&	100$\pm$10		\\
J16090075--1908526	&	39.8	&	36.1	&	3763	$\pm$	26	&	1.20	$\pm$	0.08	&	--10.88	$\pm$	0.12	&	0.125		$\pm$	0.008	&	1.3	&	1.2$\pm$0.1	&	3.1$\pm$2		\\
J16120668--3010270	&	10.9	&	9.8	&	3988	$\pm$	26	&	1.07	$\pm$	0.03	&	--10.54	$\pm$	0.11	&	0.970		$\pm$	0.057	&	1.3	&	17$\pm$1 &	5.4$\pm$0.3		\\
J16124373-3815031	&	18.4	&	11.2	&	3769	$\pm$	28	&	1.25	$\pm$	0.05	&	--10.43	$\pm$	0.12	&	0.048		$\pm$	0.002	&	1.4	&	0.71$\pm$0.04	&	3.5$\pm$0.2		\\
RX\, J1615.3--3255	&	26.4	&	32.4	&	4317	$\pm$	26	&	1.71$\pm$	0.08	&	--9.96	$\pm$	0.11	&		0.345	$\pm$	0.020	&	5.5	&	8.5$\pm$0.5	&	48$\pm$2		\\
J16202863--2442087	&	42.6	&	21.3	&	3593	$\pm$	29	&	1.11	$\pm$	0.06	&	--11.91	$\pm$	0.12	&	0.074		$\pm$	0.004	&	1.1	&	4.7$\pm$0.3	&	6.4$\pm$0.2		\\
AS 209	&	72.0	&	38.3	&	4346	$\pm$	29	&	2.22	$\pm$	0.12	&	--8.25	$\pm$	0.12	&	0.255		$\pm$	0.014	&	3.3	&	3.8$\pm$0.2	&	26$\pm$1		\\
WRAY15--1880	&	22.6	&	22.1	&	4420	$\pm$	29	&	1.67	$\pm$	0.08	&	--8.19	$\pm$	0.10	&	0.073		$\pm$	0.005	&	19	&	31$\pm$2	&	30$\pm$2		\\
WRAY15--1880	&	24.8	&	19.0	&	4310	$\pm$	26	&	1.65	$\pm$	0.09	&	--8.12	$\pm$	0.08	&	0.062		$\pm$	0.003	&	23	&	35$\pm$2	&	16$\pm$1		\\
RX\, J1852.3-3700	&	50.3	&	67.5	&	4253	$\pm$	28	&	1.35	$\pm$	0.07	&	--9.59	$\pm$	0.11	&		2.841	$\pm$	0.158	&	8.1	&	220$\pm$10	&	22$\pm$1		\\
\hline
\end{tabular}
\tablefoot{$^a$ $a_{\rm min}$(inner) corresponds to the value where temperature of grains is equal to the sublimation temperature of silicates. }
\label{tab:output_data}
\end{table*}

\begin{table*}[h]
\caption{Derived quantities.}
\centering
\tiny
\begin{tabular}{lcccccccl}
\hline
\hline
Star	&$R_{\rm max,grain,inner}$	 & $\beta$ &$M_{\rm dust,inner}$	&	$M_{\rm dust,outer}$	&	Age	&	Age lit. & Ref.&Comments	\\
& (cm) && (\MEarth) & (\MEarth) & (Myr)& (Myr)& \\
\hline
BP Tau	&	2.0	$\pm$	0.4	&--2.781$\pm$0.008&	0.018	&	26	&	3.5	$\pm$	0.6	&	1.8	&	12	&		\\
IP Tau	&	0.097	$\pm$	0.001	&--2.872$\pm$0.008&	0.0036	&	4.3	&	3.2	$\pm$	0.6	&	5.1	&	14	&	Ring	\\
IQ Tau	&	2.7	$\pm$	0.5	&--2.842$\pm$0.008&	0.018	&	41	&	1.5	$\pm$	0.2	&	2.9	&	3	&		\\
GK Tau	&	3.0	$\pm$	0.5	&--2.817$\pm$0.008&	0.062	&	13	&	1.7	$\pm$	0.3	&	2.9	&	4	&	Central peak	\\
CI Tau	&	3.0	$\pm$	0.5	&--2.810$\pm$0.008&	0.037	&	38	&	2.5	$\pm$	0.4	&	1.7	&	10	&	Central peak; possible dust trap	\\
LkCa 15	&	0.0010	$\pm$	0.0001	&--2.980$\pm$0.008&	0.00002	&	92	&	2.8	$\pm$	0.3	&	1.7	&	10	&		\\
DR Tau	&	2.0	$\pm$	0.4	&--2.804$\pm$0.008 &0.35	&	83	&	0.7	$\pm$	0.1	&	0.6	&	2	&	Central peak	\\
DS Tau	&	1.0	$\pm$	0.2	&--2.836$\pm$0.008&	0.0056	&	16	&	3.0	$\pm$	0.4	&	1.9	&	7	&	Central peak	\\
GM Aur	&	0.013	$\pm$	0.002	&--2.897$\pm$0.008&	0.0014	&	51	&	2.2	$\pm$	0.4	&	1.9	&	7	&	Rings	\\
V836 Tau	&	0.13	$\pm$	0.03	&--2.870$\pm$0.007&	0.00056	&	38	&	2.7	$\pm$	0.4	&	3.1	&	9	&	Central peak	\\
MWC 758	&	0.016	$\pm$	0.004	&--2.940$\pm$0.009&	0.0025	&	24	&	2.6	$\pm$	0.6	&	3.5	&	6	&	20	\\
CQ Tau	&	0.010	$\pm$	0.002	&--2.938$\pm$0.009&	0.0026 &	71	&	2.4	$\pm$	0.4	&	8.9	&	13	&	22	\\
SY Cha	&	0.023	$\pm$	0.004	&--2.941$\pm$0.008&	0.00034	&	116	&	2.4	$\pm$	0.4	&	1.5	&	8	&		\\
PDS 66	&	2.3	$\pm$	0.4	&--2.810$\pm$0.008&	0.017	&	25	&	4.3	$\pm$	0.8	&	8.5	&	15	&	Central peak	\\
PDS 70	&	0.028	$\pm$	0.005	&--2.936$\pm$0.005&	0.00016	&	30	&	5.5	$\pm$	0.8	&	5.4	&	16	&	Spiral	\\
Sz 65	&	3.0	$\pm$	0.5	&--2.800$\pm$0.009&	0.012	&	17	&	1.5	$\pm$	0.2	&	1.9	&	1	&	Central peak	\\
GW Lup	&	1.3	$\pm$	0.2	&--2.770$\pm$0.008&	0.028	&	34	&	2.4	$\pm$	0.4	&	2.0	&	1	&	Central peak	\\
GQ Lup	&	3.0	$\pm$	0.5	&--2.810$\pm$0.008&	0.050	&	56	&	1.0	$\pm$	0.2	&	0.9	&	1	&	Central peak	\\
Sz 77	&	2.1	$\pm$	0.4	&--2.785$\pm$0.008&	0.011	&	19	&	1.7	$\pm$	0.2	&	3.0	&	1	&	Central peak	\\
RX\, J1556.1--3655	&	1.0	$\pm$	0.2	&--2.844$\pm$0.008&	0.0053	&	25	&	3.2	$\pm$	0.5	&	...	&	...	&		\\
IM Lup	&	3.0	$\pm$	0.5	&--2.810$\pm$0.008&	0.033	&	183	&	0.9	$\pm$	0.1	&	0.5	&	1	&	Central peak	\\
RU Lup	&	3.0	$\pm$	0.5	&--2.810$\pm$0.007&	0.59	&	48	&	1.6	$\pm$	0.3	&	0.8	&	1	&	Central peak	\\
Sz 129	&	0.010	$\pm$	0.002	&--2.918$\pm$0.008&	0.00001	&	22	&	4.2	$\pm$	0.7	&	3.5	&	1	&		\\
RX\, J1604.3--2130A	&	0.010	$\pm$	0.002	&--2.910$\pm$0.009&	0.00003	&	85	&	8	$\pm$	4	&	11.3	&	19	&	Ring	\\
V1094 Sco	&	1.0	$\pm$	0.2	&--2.834$\pm$0.008&	0.0025	&	332	&	6.3	$\pm$	1.1	&	2.5	&	18	&	Central peak	\\
2MASS J16090075--1908526	&	3.0	$\pm$	0.5	&--2.811$\pm$0.008&	0.012	&	10	&	8	$\pm$	2	&	...	&	...	&	Ring	\\
2MASS J16120668--3010270	&	0.35	$\pm$	0.05	&--2.857$\pm$0.007&	0.015	&	18	&	9	$\pm$	1	&	3.5	&	17	&	Ring	\\
2MASS J16124373-3815031	&	1.0	$\pm$	0.2	&--2.844$\pm$0.008&	0.0024	&	11	&	5.7	$\pm$	0.8	&	...	&	...	&	Central peak	\\
RX\, J1615.3--3255	&	0.07	$\pm$	0.01	&--2.901$\pm$0.008&	0.00050	&	158	&	2.7	$\pm$	0.4	&	3.0	&	...	&	21	\\
2MASS J16202863--2442087	&	0.008	$\pm$	0.001	&--2.942$\pm$0.008&	0.00005	&	21	&	7.5	$\pm$	1.1	&	...	&	...	&		\\
AS 209	&	3.0	$\pm$	0.5	&--2.800$\pm$0.008&	0.40	&	85	&	1.7	$\pm$	0.2	&	0.8	&	5	&	Central peak	\\
WRAY15--1880	&	0.016	$\pm$	0.002	&--2.868$\pm$0.008&	0.00011	&	98	&	3.3	$\pm$	0.5	&	2.8	&	11	&		\\
WRAY15--1880	&	0.011	$\pm$	0.002	&--2.867$\pm$0.009&	0.00005	&	51	&	3.4	$\pm$	0.6	&	2.8	&	11	&	Weak asymmetries	\\
RX\, J1852.3-3700	&	0.0037	$\pm$	0.0006	&--2.945$\pm$0.009&	0.00019	&	72	&	5.6	$\pm$	0.9	&	6.0	&	11	&	Weak asymmetries	\\
\hline
\end{tabular}
\tablebib{
(1) \citet{Frasca2017};
(2) Age of L1558 (Gratton et al, in preparation);
(3) Age of L1529 (Gratton et al, in preparation);
(4) Age of L1524 (Gratton et al, in preparation);
(5) \citet{Fedele2018};
(6) \citet{Meeus2012};
(7) Age of L1517 (Gratton et al, in preparation);
(8) \citet{Galli2021};
(9) Age of L1544 (Gratton et al, in preparation);
(10) Age of L1536 (Gratton et al, in preparation);
(11) \citet{Rigliaco2025};
(12) Age of L1495/B209 (Gratton et al, in preparation);
(13) \citet{Vioque2018};
(14) Age of HD28354 (Gratton et al, in preparation);
(15) \citet{Ribas2023};
(16) \citet{Keppler2019};
(17) \citet{Trapman2025};
(18) \citet{vanTerwisga2018};
(19) \citet{Pecaut2013};
(20) Strong asymmetric features \citep{Baruteau2019}. 21. Peak in ALMA image at a separation of 16 mas (2.5 au) and $PA=211$ degree that can be around the secondary; if this the position of the secondary, its mass should be about 0.09 \Msun. Age is then from the radius for the  mass of the primary taking into account this secondary. 22. Strong asymmetric features \citep{Uyama2020, Wolfer2021}.
}
\label{tab:derived_data}
\end{table*}


\section{Density equations}
\label{Sect:equations}
\relax 

In this Appendix we describe the equations used to relate the gas and dust surface densities to the total mass of the inner and outer discs.

In general, the mass $M_{\rm disc}$\ of a centrally symmetric flat disc may be written as:
\begin{equation}
M_{\rm disc} = \pi \int_{a_{\rm min}}^{a_{\rm max}} \sigma(a)\, a^2 da
\end{equation}
where $\sigma(a)$ is the surface density as a function of the radial distance $a$, and $a_{\rm min}$ and $a_{\rm max}$ are the minimum and maximum radii of the disc. We assume that $\sigma(a) \sim 1/a$ in the inner disc (which is generally unresolved in ALMA images), while $\sigma(a) \sim I(a)$ in the outer disc, where $I(a)$ is the run of the intensity of the continuum emission along the major axis as measured in band 6 or 7 by ALMA. 

We further assumed that the disc is composed of dust and gas, so that we may write:
\begin{equation}
\sigma(a) = \sigma_D(a)+\sigma_G(a),
\end{equation}
where $\sigma_D(a)$\ and $\sigma_G(a)$\ are the surface density of gas and dust, respectively. In general, $\sigma_G(a)>>\sigma_D(a)$. The fractional mass in dust $f(a)$\ is then:
\begin{equation}
f(a) = \frac{\sigma_D(a)}{\sigma_D(a)+\sigma_G(a)}.
\end{equation}
We further assume that dust is the combination of two different types of grains, silicate and ice grains with (maximum) fractional masses of 0.00314 and 0.00686. However, we further assume that the grains are only present in the regions where the grain equilibrium temperature is below the sublimation value that we assume to be 1500 and 170 K for silicates and ices, respectively. Let us call $a_{\rm sil}$ and $a_{\rm ice}$ the radii that limit these regions. The maximum fractional mass of grains $f(a)_{\rm max}$ is 0.00314 for $a_{\rm sil}<a<a_{\rm ice}$ and 0.01 for $a>a_{\rm ice}$, that is, gas-to-dust ratios of 318 and 100, respectively. In this approximation, $f(a)_{\rm max}=0$ for $a<a_{\rm sil}$. Since in our model there is no (continuum) emission from this region, we assumed that $a_{\rm min}\geq a_{\rm sil}$, although there should be some gas in this very inner region to sustain the accretion on the star.

We further assume that the grains have a distribution of radii $r$\ described by a power law with exponent $\beta$. Grain radii are needed to compute the thermal emission (see Sect. \ref{sect:methods}). After several trials, we found that a value of $\beta=-2.81$ matches data. In our model, the actual fractional mass $f_a$\ in dust at a given distance $a$ from the star is then:
\begin{equation}
\frac{f(a)}{f(a)_{\rm max}} = \frac{\int_{R_{\rm min}}^{R_{\rm max\, disc}} r^\beta \, r^3 dr}{\int_{R_{\rm min}}^{R_{\rm max}} r^\beta \, r^3 dr},
\end{equation}
where $R_{\rm min}$ and $R_{\rm max}$ are the minimum and maximum size of grains. We always assumed $R_{\rm min}=0.1$\, $\mu$m. For the outer disc, we assumed $R_{\rm max\, disc}=R_{\rm max}=3$\, cm (so that $f_a=f(a)_{\rm max}=0.01$), but we left this quantity free in the inner disc, although we forced it to be $R_{\rm max\,  disc}\leq R_{\rm max}$. For the inner disc $f_a\leq f(a)_{\rm max}$, meaning that the gas-to-dust ratio in the inner disc may be higher than 318, sometimes by a large factor, depending on the value of $R_{\rm max,grain,inner}$, that is a free parameter in our model.

\end{appendix}

\end{document}